\documentclass{article}

\usepackage{arxiv}

\usepackage{CJKutf8} 
\usepackage[utf8]{inputenc} 
\usepackage[T1]{fontenc}    
\usepackage{hyperref}       
\usepackage{url}            
\usepackage{booktabs}       
\usepackage{amsfonts}       
\usepackage{nicefrac}       
\usepackage{microtype}      
\usepackage{lipsum}
\usepackage{graphicx}
\usepackage{enumitem} 
\usepackage{threeparttable}
\usepackage{colortbl}
\usepackage{tcolorbox}
\usepackage{xcolor, soul}
\graphicspath{ {./images/} }

\title{Cash, Credibility, and Conversion: The Influence of Synthetic Media on Investment Behavior}

\author{
 Nicholas G. Keeley  \\
  Schwarzman College\\
  Tsinghua University\\
  Beijing, China\\  \texttt{nicholas.keeley@sc.tsinghua.edu.cn} \\
    Advised by Dr. Zeng Yi, Chinese Academy of Sciences\\
  }

\begin{document}
\maketitle
\begin{abstract}
Prior to November of 2022, the topic of synthetic media was largely buried within academic journals, constrained to conversations about national security, and often fundamentally misunderstood. The release of ChatGPT, however, has accelerated discourse on the societal impacts of synthetic media. This study first highlights several gaps within existing literature on synthetic media, structuring the impact potential and limitations of synthetic media threats within a theoretical framework. Second, it identifies financial information environments as prime candidates for future disruption via synthetic text modalities, proposing an experimental survey for measuring the influential power of synthetic financial text on global investment communities. Rather than merely assessing the ability of survey participants to distinguish genuine from synthetic text, the experiment contained within this study measures synthetic media influence by observing its ability to manipulate \textit{belief} via a series of behavioral variables. The results indicate that synthetic text can significantly shift investor sentiment away from what it might otherwise have been under truthful information conditions. Furthermore, synthetic financial text demonstrated a unique ability to ``convert" investors, inspiring \textit{extreme} changes in outlook about a company compared to genuine financial texts. This trend should inspire concern within the global financial community, particularly given the historical vulnerability of equity markets to investor sentiment shocks. \footnote{Please send correspondence to \texttt{ngkeeley8@gmail.com}. In order to meet preprint specifications, this paper was adapted from its original format, as submitted to Tsinghua University in partial fulfillment of the requirement for the degree of Master of Management Science in Global Affairs.} 
\end{abstract}
\twocolumn


\section{Introduction}
\subsection{The Significance of Two Events, One World Apart}

In November of 2022, two seemingly disconnected events on opposite ends of the world brushed paths in time. One generated \$450 billion USD, while another revolutionized modern technology. But together, they may foreshadow a dark future for the global financial world. 

The first of these events was triggered by the dissemination of an anonymous screenshot on China’s most prominent social media application, WeChat \begin{CJK*}{UTF8}{gbsn}微信\end{CJK*}. The image contained a piece of paper with four paragraphs of Chinese characters, detailing the intended loosening of China’s Covid Zero policy. Nobody verified the image's authenticity or identified an original author, yet the image managed to spread from WeChat, to investment forums on Twitter, and ultimately to Chinese trading platform, Xueqiu \begin{CJK*}{UTF8}{gbsn}雪球\end{CJK*} (``How a Mysterious China Screenshot Spurred \$450 Billion Rally,” 2022). Although Chinese officials did not officially repeal the Covid Zero policy until over a month later (Zhu et al., 2023), the image inspired enough trading activity to shift equity markets by almost half a trillion dollars.

\begin{figure}
  \centering
  \includegraphics[width=0.8\linewidth, height=0.8\linewidth]{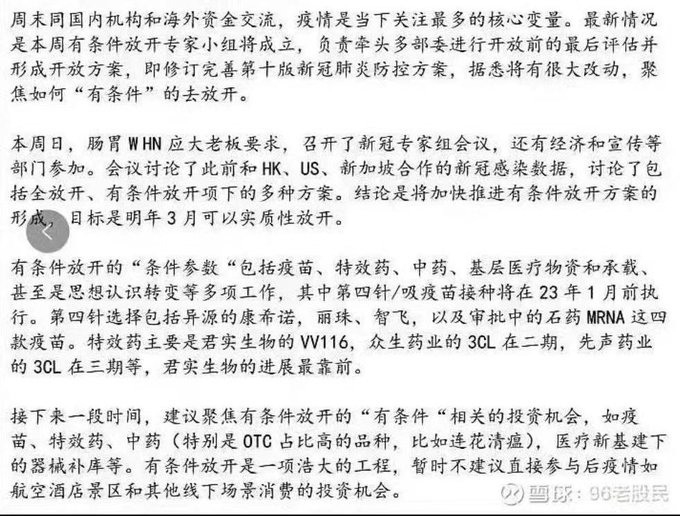}
  \caption{Anonymous image of China's Covid Zero reopening on Twitter (“How a Mysterious China Screenshot Spurred \$450 Billion Rally,” 2022).}
  \label{fig:bloomberg_photo}
\end{figure}

The second of these events occurred in America: the release of OpenAI’s ChatGPT. Since then, the large language model (LLM) chatbot has become ``the fastest-growing consumer application in history,” accruing an average user count of 13 million per month (Hu, 2023).  While its users have developed countless innovative use-cases for ChatGPT, researchers have grown increasingly concerned about its ability to spread misinformation online (Hsu \& Thompson, 2023).

This study examines the impact of synthetic media on society, and while ChatGPT and the anonymous text passage described above were released separately, the possibility of their convergence portends a frightening reality for the financial world which may rapidly be coming to fruition. The use of these two events also highlights the significance of synthetic media as an area of study, as well as several characteristics that render it more nuanced than it might initially appear.

 First, synthetic content generation lies at the confluence of several expansive subject domains, including cutting-edge machine learning techniques (e.g. LLMs), epistemological questions about authenticity and credibility, and even ethical considerations (e.g. free speech protections).
 
 Second, the examples reaffirm that this is a global problem, not a localized one. Language models developed in one country can be used to generate content in another, and the financial effects that result can yield economic impacts that transcend international boundaries. 
 
 Third, the anonymous screenshot described above was not just an image, but an image of \textit{text}. While this study focuses on synthetic text, the anonymous screenshot serves as a reminder that synthetic media has multiple modalities; it can consist of videos, images, audio files, text, or even a combination thereof. 

 Finally, the positive and negative externalities of both events underscore that the moral value judgments underlying synthetic media are difficult to disentangle. Indeed, this paper deliberately employs the phrase “synthetic media”\footnote{In reality, this phrase has many names within existing literature, including “deepfakes” (Helmus, 2022), “AI-generated content” (Cao et al., 2023), “machine generated” content (Crothers et al., 2023) “deep synthesis information content” \begin{CJK*}{UTF8}{gbsn} 深度合成信息内容 (《互联网信息服务深度合成管理规定》答记者问-中共中央网络安全和信息化委员会办公室\end{CJK*}, 2022), etc. } (Bateman, 2020) instead of “deepfake” to avoid negative biases embedded within preexisting (primarily English) literature, to prevent the conflation of truth value with provenance integrity, to acknowledge the subjectivity separating media augmentation from media manipulation, and to encompass the many methods of delivery and modalities that synthetic media can take. Furthermore, the term “media” implies content that is intended to impart information, while the term “synthetic” underscores a difference with manually-produced media – a distinction described in more detail in Chapter 2. 

 \subsection{Research Questions}

 The primary research question of this study was to determine if ChatGPT generates synthetic financial disinformation that is more persuasive than genuine financial information:
 

\begin{itemize}
    \item \textbf{H1}: GPT-3.5\footnote{As described later in more detail, GPT-3.5 is the model upon which ChatGPT was originally built and corresponds with the version of ChatGPT used for this study's experimental survey.} produces \textit{synthetic} financial disinformation that is more persuasive than \textit{genuine} financial information.
\end{itemize}

\noindent In addition to statistically testing the main hypothesis, this study conducted exploratory analysis of several sub-hypotheses in parallel:\newline

\begin{itemize}
    \item \textbf{H2}: GPT-3.5 produces synthetic \textit{financial} disinformation that has more persuasive power than synthetic \textit{world news} disinformation.\newline
    
    \item \textbf{H3}: Synthetic financial disinformation about \textit{niche industries} is more persuasive than that of \textit{non-niche industries}.
\end{itemize}
\section{Literature Review}
\subsection{Defining Synthetic Media: Avoiding Detection By Humans and Machines}

The history of media alteration dates back to the origins of photography (Verdoliva, 2020; Digital Forensics, 2008; Chesney \& Citron, 2019). For example, the first permanent photograph in history was taken in 1828 (Digital Forensics, 2008). Less than forty years later, a photograph of President Abraham Lincoln emerged which had been crafted by splicing portions of two different images. Doctoring convincing images has since grown more accessible and effective via technology applications, such as Adobe’s suite of ``Photoshop” tools (Verdoliva, 2020; Chesney \& Citron, 2019). This kind of traditional media alteration can be highly convincing if carried out professionally (Verdoliva, 2020). However, it is distinct in that it is manual in nature. Conversely, the term ``synthetic media” refers to images, videos, audio files, and text that are entirely generated via artificial intelligence (Bateman, 2020). When produced correctly, synthetic media can be so convincing that it deceives not only the naked eye (Helmus, 2022), but also advanced detection algorithms (Deepfake Detection Challenge Results, 2020) - which may otherwise have served as societal safety nets for human bias and error. 

For example, multiple empirical studies demonstrate that humans perform no better than guessing when they are distinguishing between synthetic and genuine videos, images, audio files, and text (Bray et al., 2022; Korshunov \& Marcel, 2020; Mai et al., 2023; Brown et al., 2020). In fact, one study suggests that human subjects mislabel over 75\% of high-quality synthetic videos as genuine footage (Korshunov \& Marcel, 2020).\footnote{Another study found these results to be robust to language differences - specifically Chinese and English (Mai et al., 2023)} Unfortunately, detection algorithms used by organizations to sift through disinformation at-scale (Morrish, 2023; Deepfake Detection Challenge Results, 2020) have also struggled to reliably classify synthetic media. As a prime example, social media giant Meta partnered with Microsoft and multiple global universities in 2020 to host the ``Deepfake Detection Challenge” (DFDC). Over 2,000 applicants submitted 35,000 detection models. Yet the highest performing model only achieved an accuracy rating of 65.18\% on ``real world” synthetic video detection (Deepfake Detection Challenge Results, 2020).  Similar results have been demonstrated empirically for detecting synthetic images, audio files, and text (Du et al., 2020; Müller et al., 2022; Pu et al., 2022). 

The ability of synthetic media to bypass both human \textit{and} algorithmic detection distinguishes it from manually-altered media, and is a distinction worth underscoring again, considering the ease with which it can be overlooked. The reason behind this comparative advantage is better understood through the evolutionary history of synthetic media creation algorithms, and the ``generalization problem" that has accompanied it.

\subsection{The Evolution of Synthetic Media Creation and the ``Generalization Problem"}

One of the more commonly used terms in preexisting English literature on synthetic media is the term “deepfake.” It was coined by a Reddit user in 2017 (Somers, 2020) who developed a technique for superimposing the faces of celebrities onto pornographic videos using “deep” machine learning models called “neural network[s]” (Chesney \& Citron, 2019). Broadly speaking, neural networks are statistical models that mimic the synaptic architecture of the human brain to form predictions (Chesney \& Citron, 2019). The earliest deepfakes were formed by combining one neural network’s inputs (trained to translate celebrity faces into an embedding vector) with another neural network’s outputs (trained to translate embedding vectors into pornographic videos) (Nguyen et al., 2022; BM, 2022).

Since 2017, methodological advancements to the neural architectures underlying deepfakes have significantly amplified the quality of synthetic media products – as well as their ability to elude detection. For audio-visual modalities, one of the first major milestones included the application of “Generative Adversarial Networks” (GANs). Developed by Google researcher Ian Goodfellow, GANs pit a “forger” neural network against a “detector” neural network in a prolonged competition, ultimately generating a highly convincing synthetic product that can deceive detection algorithms (Helmus, 2022; Chesney \& Citron, 2019). Additional methods within the synthetic audio-visual domain include Variational Auto-Encoders (VAEs), as well as generative diffusion models, introduced by Stability.AI in 2022 (Cao et al., 2023). 

From a synthetic text perspective, the first adaptation of neural network architecture was the introduction of Recurrent Neural Networks (RNN), allowing for the use of longer input phrases (Cao et al., 2023). Next came the development of mechanisms for incorporating “memory” into a network's statistical weighting process, including Long Short-Term Memory (LSTM) and Gated Recurrent Units (GRU) (Cao et al., 2023). However, arguably one of the most important milestones in the history of synthetic text development was the adaptation and use of the “Transformer” architecture, proposed by Vaswani et al. (Cao et al., 2023; Crothers et al., 2023; Vaswani et al., 2017). The Transformer architecture, currently hailed as “the state-of-the-art in natural language generation across natural language tasks” (Crothers et al., 2023), incorporates a “multi-head attention” mechanism that allows models to efficiently focus on the most important portions of a text input and output (Crothers et al., 2023; Cao et al., 2023; Brown et al., 2020; Ruby, 2023). Consequently, an “overwhelming majority” (Crothers et al., 2023) of modern generative language models have adopted it – including the one underlying ChatGPT.\footnote{In fact, audio-visual models have also adopted Transformer architectures, such as Swin Transformer and Vision Transformer. The use of this architecture across models of differing modalities has also enabled the creation of “multimodal models” (e.g. CLIP and DALL-E-2), which can perform cross-modality tasks such as “text-to-image generation” (Cao et al., 2023).} 

ChatGPT, which was released by OpenAI in November of 2022 (Ruby, 2023), allows users to type chatbot-style questions and receive immediate feedback. Its interface was technically built upon OpenAI’s “GPT-3.5” series of language models; however, these models are merely slight improvements on GPT-3 – their “core” model (Introducing ChatGPT, 2022; OpenAI, n.d.; New GPT-3 Capabilities, 2022).\footnote{GPT stands for “Generative Pre-trained Transformer” (Crothers et al., 2023). Since this publication, ChatGPT Plus – ChatGPT's paid version – has also begun accepting inputs from its newest model: GPT-4 (OpenAI, 2023).} Two factors distinguished GPT-3 from its predecessor, which possessed an almost identical architectural framework. First, GPT-3’s architecture grew exponentially in size – from 1.5 billion parameters (GPT-2), to approximately 175 billion parameters (Cao et al., 2023). Second, the data used for model pre-training grew from 38 gigabytes (GPT-2), to 570 gigabytes (Cao et al., 2023). This jump in scale, the model’s creators found, led to rapid increases in the efficient handling of “in-context information” (Brown et al., 2020). Coupled with human-augmented reinforcement learning (Cao et al., 2023) and a free, user-friendly interface, ChatGPT seized news headlines for months following its release as users found innovative – as well as concerning – applications of its technology (Roose, 2022; Huang, 2023).

The continuous development of new generative models and “adversarial [techniques] that fool detection algorithms” (Crothers et al., 2023) described above has surfaced a significant obstacle for automated synthetic media detection. Referred to here as ``the generalization problem," this phenomenon describes the tendency of detection algorithms to struggle while labeling real-world synthetic content created by newly emergent models and methods (Deepfake Detection Challenge Results, 2020; Du et al., 2020; Müller et al., 2022; Pu et al., 2022).\footnote{One anecdote captures this phenomenon in practice particularly clearly. In 2018, researchers openly published a paper identifying subject blink rates as a critical feature for distinguishing synthetic from genuine videos. Within several weeks, new synthetic media emerged from malign algorithms specifically adapted to incorporate realistic blink rates (Helmus, 2022).} In this condition, detection algorithms can be described as “fail[ing] to generalize” to real-world data (Pu et al., 2022; Deepfake Detection Challenge Results, 2020; ). For example, one study on falsified audio-file detection reported “[model] performance degradation of up to one thousand percent” when exposed to new synthetic content (Müller et al., 2022). Even OpenAI’s most recent synthetic text detection algorithm only accurately classified 26\% of the data it was evaluated on (Kirchner et al., 2023). The generalization problem has locked synthetic media generators and detectors into a perpetual technology “arms race” (Helmus, 2022). As a result, synthetic media detection via algorithmic means is, as described in Meta’s DFDC results summary, “still very much an unsolved problem” (Deepfake Detection Challenge Results, 2020). 

\subsection{A History of Synthetic Media Use: Trending Towards Politics}

While understanding the evolution of synthetic content generation reveals important insights about the advantages and limits of their \textit{implementation}, examining trends across publicized synthetic media use-cases sheds light on the \textit{objectives} behind their use. The section that follows examines both socially beneficial and deliberately malicious synthetic media use-cases that are either mentioned in preexisting literature, or sensationalized in the news. For simplicity, it does so by dividing the four synthetic media modalities into two categories: audio-visual (video, image, audio), and text. Of special note to later analysis, a dominant proportion of malicious use-cases, across both categories, tend to involve politics.

\subsubsection{Audio-Visual Use-Cases}

Perhaps the most commonly cited synthetic media products on the internet today are satirical videos mimicking former President Obama (Vincent, 2018; BuzzFeedVideo, 2018) and actor Tom Cruise (Helmus, 2022; Vecanoi, 2021). These videos have garnered millions of views online. However, synthetic media applications have also extended to use-cases with higher stakes. For instance, political candidates in both India and Korea have leveraged synthetic media to amplify the effectiveness of their digital campaigns. In 2020, Indian parliamentary candidate Manoj Tiwari “used a ‘lip-synch’ deepfake algorithm” (Lyons, 2020) to "speak” a dialect he couldn’t naturally speak. The tactic allowed him to reach 15 million Indian voters “he might not have otherwise been able to reach” (Lyons, 2020). Two years later, advisors to South Korean presidential candidate (now current president) Yoon Suk-yeol published campaign videos using his more youthful “deepfake avatar,” (Shin \& Yi, 2022) raising public opinion ratings for young adults by 10\% (Shin \& Yi, 2022).

\begin{figure}
  \centering
  \includegraphics[width=0.8\linewidth]{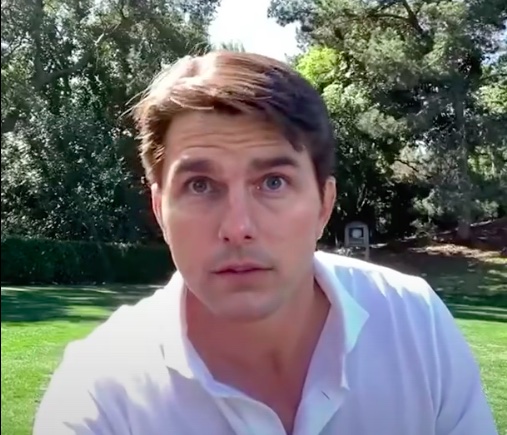} 
  \caption{Screenshot of Tom Cruise ``deepfake" (Vecanoi, 2021).}
  \label{fig:tc}
\end{figure}

Multiple international companies have also commercialized synthetic media to serve their consumers. Most notably in China, technology firm Momo Inc. \begin{CJK*}{UTF8}{gbsn}陌陌技术有限公司\end{CJK*} released the ZAO mobile application in September of 2019, allowing users to superimpose their faces onto the images of celebrities (Horwitz, 2019). During its weekend release, millions of downloads led to the application becoming the most popular free application on China’s iPhone App Store (Horwitz, 2019). Though the application is no longer made available by Apple due to user privacy complaints (Horwitz, 2019) and an intervention by China’s Ministry of Industry and Information Technology (Regulating Tech Use to Curb “deepfake” Risks, 2022), ZAO’s service has been replicated in multiple other companies since, to include Russia’s FaceApp (Stokel-Walker, 2019), America’s Snapchat (Nuñez, 2020), and China’s TikTok (Nuñez, 2020).  

More recent studies suggest that these same features may be capable of fulfilling consumer needs in the healthcare sector, too. For instance, sexual assault survivors have donned virtual reality headsets to undergo trial exposure therapy in synthetically generated digital environments (Wiederhold, 2021; van Minnen et al., 2022).

Unfortunately, the urgency within existing literature for detecting synthetic media typically stems from more insidious use-cases that are increasingly frequent – not to mention damaging (Hwang, 2020; Helmus, 2022; Chesney \& Citron, 2019; Sensity, 2019). For example, according to a report generated by Sensity (formerly ``Deeptrace”), the number of synthetically produced videos online increased from almost 8,000 to 14,678 between 2018 and 2019. Of these videos, an overwhelming proportion represented pornographic content, garnering over 13 million views (Sensity, 2019). While the vast majority of subjects in these videos were high-profile, international celebrities, 100\% of the videos targeted women – revealing reputational damage severely concentrated along gender lines (Chesney \& Citron, 2019; Sensity, 2019).   

Fraud and theft have also been attributed to the malicious use of synthetic media. In 2019, thieves used algorithmically altered voiceovers to convince the Chief Executive Officer (CEO) of a British business to transfer \$200,000 to fake accounts (Damiani, 2019). Two years later, another businessman mistakenly transferred over \$35 million to fraudsters via a similar scheme (Brewster, 2021). 

However, manipulation of political information environments – both domestically, and internationally – represents the largest focal point of preexisting literature on synthetic, audio-visual media. According to the “AI, Algorithmic and Automation Incidents and Controversies” (AIAAIC) repository, political content accounted for 34.7\% of audio-visual deepfakes since 2018 – the single largest category other than “Media/entertainment/sports/arts” at 40.3\% (AIAAIC - AI and Algorithmic Incidents and Controversies, 2019).\footnote{Calculated by filtering “Sector(s)” column for entries containing the term “Politics” or “Govt.”} The most recent – and perhaps shocking – example occurred in March of 2022 (AI Index Report 2023 – Artificial Intelligence Index, 2023; Simonite, 2022). Less than a month after Russian forces invaded Ukraine, a video of President Vladimir Zelensky standing wearily behind a podium emerged on Ukrainian national media outlets. In the video, he “call[ed] on his soldiers to lay down their weapons,” formally announcing the surrender of his country (Mikael Thalen [\@MikaelThalen], 2022). Of course, the video was later determined as a fake, generated to mimic Zelensky's facial structure, tone of voice, and even eye movements (Simonite, 2022; Wakefield, 2022; Allyn, 2022). While international social media channels fortunately reacted quickly in discrediting the video as a Russian information warfare tactic (Academic Webinar, 2022), its publication demarcates one of the first documented, international “weaponizations” of synthetic media in modern history (Simonite, 2022). 

\begin{figure}
  \centering
  \includegraphics[width=0.8\linewidth]{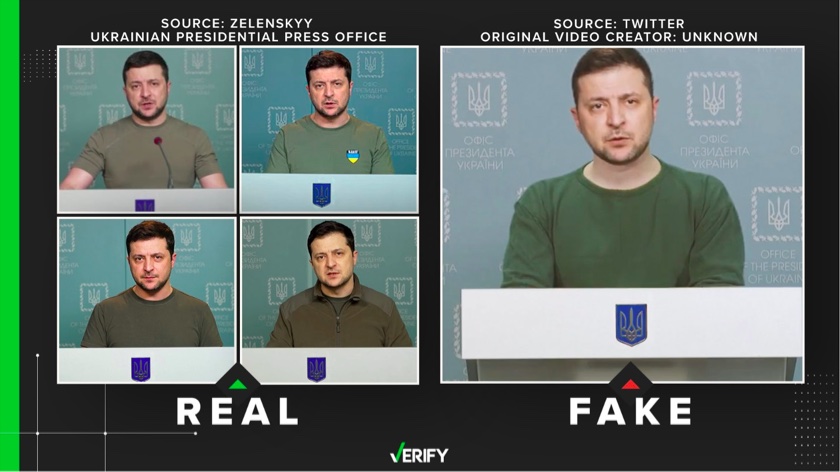} 
  \caption{Genuine President Zelensky photos vs. deepfake screenshot (AI Index Report 2023 – Artificial Intelligence Index, 2023).}
  \label{fig:zelensky}
\end{figure}

Politically motivated synthetic videos have emerged elsewhere in the world, too. For instance, in 2018, allegations about a government-sponsored “deepfake” fueled a domestic coup in Gabon (Transcript - Why the Zelensky Deepfake Failed, 2022). Although the coup was suppressed, the incumbent administration never released evidence denying claims of synthetic media creation, and the video’s authenticity remains in question to this day (Transcript - Why the Zelensky Deepfake Failed, 2022). In 2021, a similarly questionable video implicating government leader Aung San Suu Kyi in a bribery accusation surfaced in Myanmar (Gregory, 2021). Meanwhile, synthetically generated satellite images have been labeled as a potential national security threat (Zhao et al., 2021), and the specter of artificial intelligence looms over manual disinformation wars – such as the one that raged in Tigray until only recently (Ethiopia’s Warring Sides Locked in Disinformation Battle, 2021; Hiebert, 2022).  

\begin{figure}
  \centering
  \includegraphics[width=0.8\linewidth,height=0.8\linewidth]{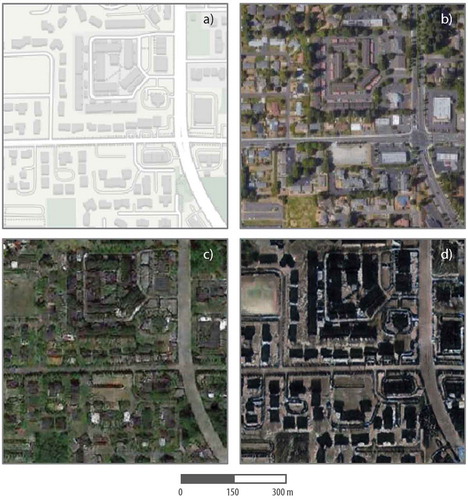} 
  \caption{(a) Original Tacoma map tile; (b) Original Tacoma satellite image; (c) Fake Tacoma image with Seattle landscape features; (d) Fake Tacoma image with Beijing landscape features (Zhao et al., 2021).}
  \label{fig:taco-bj}
\end{figure}

\subsection{Text Use-Cases}

As seen in Section 2.2, the evolutionary timeline of synthetic media generation spans less than a decade. As a result, the history of its use is short. The list of sensationalized synthetic \textit{text} use-cases is even shorter – eclipsed by concerns about audio-visual “deepfakes.”\footnote{See (Chesney \& Citron, 2019; Verdoliva, 2020; Hwang, 2020; Gregory, 2021; Helmus, 2022; Hiebert, 2022) for evidence of this imbalance.} For example, the “Deepfakes: Grounded Threat Assessment” published by Georgetown’s Center for Security and Emerging Technology (CSET) in 2020 even explicitly “eschew[ed] a close examination of ML-driven text generation” (Hwang, 2020). This may seem ironic to readers in hindsight, given ChatGPT’s current spotlight in the technology world. However, the inattention by CSET and other authors of preexisting literature underscores the speed with which LLMs and synthetically generated content have entered the public eye, as well as the lack of historical data on its use. 

The earliest origins of natural language generation began in the mid-20th century, spurred by the pursuit of language translation using machines (Jones, 2001). Following several decades of rudimentary attempts to build Q\&A-style systems, ``statistical language processing” finally became a dominant element of natural language generation in the 1990s (Jones, 2001). In 2016, a study on synthetic text detection suggested that use-cases were limited in scope from ``basic random word salads to web scraping” (Beresneva, 2016). ``The biggest part of artificial content,” it continued, ``is generated for nourishing fake web sites designed to offset search engine indexes” (Beresneva, 2016).  

Since then, the development of language models and generative techniques -- particularly the Transformer architecture -- has increased both the number and variety of synthetic media use-cases. However, like their audio-visual counterparts, malicious uses of synthetic text seem to predominantly revolve around the political domain. For example, Wired magazine wrote about a cybersecurity company’s use of GPT-2 to simulate an autonomous version of Russia’s election-tampering attempts in 2016 (Simonite, 2019; Helmus, 2022). In 2020, The Guardian rapidly (and cleverly) popularized GPT-3 by using the model to write an article about its own capabilities (“A Robot Wrote This Entire Article. Are You Scared Yet, Human?,” 2020; Helmus, 2022). GPT-2 and GPT-3 have also sparked controversy in the AI research community due to their questionable use in online forums. The former language model was once used to generate over 1,000 comments on Medicaid’s government website - indistinguishable from human comments, according to subjects in an experiment carried out in 2019 (Weiss, 2019; Crothers et al., 2023). Similarly, a model adapted from GPT-3 was used to generate over 30,000 racially and politically insensitive posts on one of 4chan’s anonymous politics threads in 2022 (Kurenkov, 2022; Crothers et al., 2023). Following a signed condemnation of the model by hundreds of signatories in the AI community (Crothers et al., 2023), the model was removed from opensource-AI hub ``Hugging Face” (Kurenkov, 2022; Crothers et al., 2023).  

In fairness, generative language models have also previously been attributed with a variety of societally beneficial tasks, including translating documents, creating text summaries, building cover letters, and serving as digital assistants (Crothers et al., 2023). As a user-friendly, publicly available tool, ChatGPT’s release has undoubtedly accelerated the proliferation of such uses cases, as well as obfuscated the boundary between “good” and “bad” uses. Indeed, ChatGPT has since been used to create poetry, comedy, essays, and even software (Roose, 2022). However, it has also raised concerns about facilitating plagiarism (Huang, 2023), introducing AI sentience (Roose, 2023), and amplifying the availability of misinformation (Hsu \& Thompson, 2023). 

\subsection{Asymmetries, Contradictions, and Gaps in Existing Literature}

Preexisting literature on synthetic media varies widely along multiple dimensions, to include technical depth, consistency, and value judgment. As alluded to in the preceding sections, this academic landscape is further complicated by asymmetrical attention to political information environments, as well as limited historical data. These characteristics can introduce contradictions and confusion about crucially relevant questions, such as evaluating the threat potential of synthetic media, quantifying its possible impact, and identifying vulnerable targets. 

For instance, less technical articles – such as ``The Supply of Disinformation Will Soon Be Infinite” (DiResta, 2020) - are often quick to identify the impending ubiquity of synthetic disinformation. However, they also often avoid explaining why this outcome is inevitable from a technological perspective (see Section 2.2 on ``the generalization problem”). Meanwhile, technical analyses often produce valuable insights on synthetic media but disagree on the threat level they pose to society. For instance, most technical studies only consider the effects of synthetic media on human subjects while overlooking their elusiveness to algorithmic detection methods (and comparative advantage over manual disinformation) (e.g. Hameleers et al., 2022; Lee \& Shin, 2022; Iacobucci et al., 2021; Shin \& Lee, 2022). Those that do manage to address both forms of permeability still differ on their definitions of threat, choosing to proxy the harms of synthetic media exposure with ``engagement intentions” (Lee \& Shin, 2022) and Turing tests (Brown et al., 2020), rather than concrete demonstrations of impact. From a more extreme perspective, several articles dismiss the additive threat of ``deepfakes” as marginal within the larger context of manual ``shallow fakes” (Helmus, 2022), while others point to more indirect erosions of trust – such as the ``liar’s dividend” (Chesney \& Citron, 2019) - as the real dangers of synthetic media.

\subsection{The Synthetic Media Development Framework}

The many moving pieces of these varied arguments cloud a methodical assessment of synthetic media’s impact on society – not to mention complicating countermeasure development. In response, this study proposes a theoretical framework that separates synthetic media development into three stages: “production,” “dissemination,” and “consumption.” Adopting this framework offers organizational structure to the complicated topics surrounding synthetic media, explains the asymmetrical focus of existing literature on political use-cases, and paves a path for addressing the more targeted research questions this study seeks to investigate.

\begin{figure}
  \centering
  \includegraphics[width=0.8\linewidth]{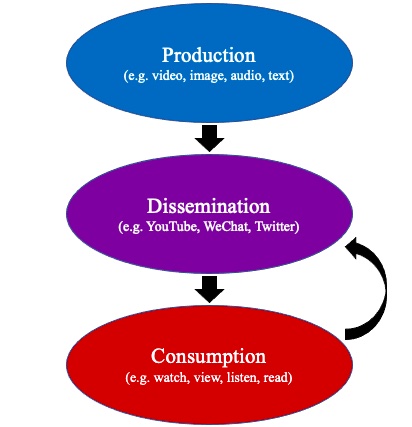} 
  \caption{The ``Synthetic Media Development Framework."}
  \label{fig:example}
\end{figure}

\subsubsection{The Production Stage}

The ``production” stage of a synthetic media product, as defined here, begins with the collection of necessary data to train a deep learning model. Because synthetic media is generated using machine learning techniques, every sub-phase of the traditional “data science pipeline” is included: data munging, model generation, testing/hyperparameter adjustment, model selection, etc. (Helmus, 2022; Data Science Pipeline, n.d.). This stage is the most time and resource-intensive stage in the framework, and is considered complete when the target synthetic media product has been generated and is ready for release. In addition to forming the foundation upon which all subsequent framework stages rest, the production stage can have a more durable effect on the synthetic media production process through the introduction of new generative algorithms and techniques.

Examining the production stage in more detail yields important insights about the threat potential of synthetic media products in the future. The first of these is the resiliency of synthetic media products to algorithmic detection, as discussed earlier in Chapter 2. However, a second and perhaps equally important characteristic of synthetic media production is how resource-intensive it is. First, a convincing synthetic media product requires enough data to train the underlying model (Hwang, 2020; Helmus, 2022). The synthetic Tom Cruise video mentioned earlier, for instance, required “hours of authentic Tom Cruise footage to train” (Helmus, 2022). Second, training and adjusting a model’s hyperparameters to produce an adequate output can require months worth of time. The model underlying the aforementioned TikTok video took over two months to train (Helmus, 2022). Indeed, the most recent natural language models for producing synthetic text required thousands of petaflop/days to train (Brown et al., 2020; Lohn \& Musser, 2022). For context, a standard laptop would take several millennia to accomplish this same task (Lohn \& Musser, 2022). Finally, this process often involves a monetary cost – either to hire data scientists or professional actors, to purchase computing power, or both. Utilizing the Tom Cruise video as a point of comparison once more, this product cost over \$10,000 USD in graphics processing units alone (Helmus, 2022).  

While critics might point to the increased ``commodification” (Hwang, 2020) of these resources through opensource data, commercial platforms, and improved computing efficiency (e.g. quantum computing) in the future, the fact remains that resource requirements currently still isolate the production of ``tailored” synthetic media (Hwang, 2020) to a narrow subsample of probable threat actors and use-cases. The extensive data requirement for training these models helps explain why existing synthetic videos are overwhelmingly targeted at celebrities and politicians, whose faces exist all over the internet via publicly available content. Meanwhile, the resource-intensive nature of synthetic media creation also helps explain why the majority of advanced, non-pornographic synthetic videos have emerged in high-reward scenarios, such as political power struggles and high-dollar heists. While a rational, resource-deficient individual or group might shoulder the cost of such an endeavor in pursuit of an urgent objective, it is reasonable to believe that they would think twice before incurring such a high cost – particularly when coupled alongside the risk of criminal attribution (Hwang, 2020). Given this information, one can reasonably expect synthetic media threats in the short term to be carried out by resource-rich organizations or individuals, and oriented towards data-available targets. 

\subsubsection{The Dissemination Stage}

The ``dissemination” stage is where content generation algorithms first exercise their comparative advantage over manual media alteration. This stage begins with a synthetic product’s release onto any media platform, and ends when the product has been seen, read, or heard by a media platform user. Presumably, the product has successfully bypassed both human and automated efforts to screen it.  

However, this stage can introduce threatening effects to society, even before consumption by individual netizens. The first of these is an ``erosion of trust in institutions,” (Chesney \& Citron, 2019) introduced by the mere existence of falsified content in the midst of genuine media (Helmus, 2022; Kreps et al., 2022). The second is an indirect consequence of the first and is called the ``liar’s dividend” (Chesney \& Citron, 2019). This phenomenon describes the use of synthetic media’s existence as a defense mechanism, debunking claims or accusations as ``fake” in order to avoid scrutiny (Chesney \& Citron, 2019). Finally, this stage interacts circularly with the consumption stage, allowing a media product – even with low levels of intrinsic credibility – to approach “viral” status (Shao et al., 2018). As consumers engage with the content, it can be reshared and even translated to other users, ultimately migrating across media platforms (Wakefield, 2022; Allyn, 2022)\footnote{This kind of migration was seen through the release of the Zelensky deepfake, in which messages were originally “amplified” (Wakefield, 2022) on Russian social media channel VKontakte before spreading to Twitter, Facebook, and YouTube.} and reaching audiences around the globe rapidly. While this paper deals exclusively with the spread of synthetic content, several studies on the spread of disinformation through synthetic actors -– such as Twitter “bots” -- are worth examining in the context of synthetically-generated media (Shao et al., 2018; Caldarelli et al., 2020; Vosoughi et al., 2018). 

\subsubsection{The Consumption Stage}

Although synthetic media that has passed through the first two stages of its development may already have yielded significant impacts on society, the “consumption” stage is particularly important because it demarcates where ideas contained as digital content transform into human \textit{action}. The consumption stage begins when a platform user interacts with synthetic media. However, the subsequent actions that succeed initial exposure are less obvious and infinitely numerous. These include providing online commentary on the product, resharing the product with others (back to the dissemination stage), trusting or dismissing its credibility, and even acting on its content (e.g. organizing a rally, investing, etc.). As seen in the use-cases above, these actions can produce effects ranging from account transfers to coups. Finally, the cyclical relationship between this stage and the dissemination stage facilitates synthetic media achieving virality and effects at-scale. Despite the crucial importance of this stage within the synthetic media development process, machine-to-human influence remains remarkably understudied. As a result, it contributes significantly to the experimental motivation described in the following section. 

\section{Experiment}

\subsection{Motive}

\subsubsection{Overlooking Belief, Impact, and Information Environments}

Examining synthetic media through a structured lens – such as the synthetic media development framework – provides a new perspective on this subject. First, the technological “arms race” (Helmus, 2022) between content generation and detection, fueled by the generalization problem of algorithmic detection, effectively guarantees the continuous introduction of innovatively manufactured media into the synthetic media development process' production stage. Unless impeded by effective countermeasures, these synthetic products will influence global information environments in potentially harmful ways as they permeate through the framework's stages.

The arrival of these products to the consumption stage yields two significant outcomes. First, it initiates a cycle of consumption and dissemination that forms the basis of content virality – a core strength of digital media (Iacobucci et al., 2021). But perhaps most importantly, it is the stage at which synthetic media crosses from the digital world as an idea, into the physical world as an action -- ultimately translating to \textit{material impact}, such as dollars lost, votes gained, and wars begun. Belief,\footnote{As shown in “Defining and Measuring Credibility Of Newspapers: Developing an Index,” the term “credibility” logically decomposes into “belief,” (Locke, 1982; Kreps et al., 2022) despite being a “multidimensional concept” (Kreps et al., 2022). Thus, these terms are used interchangeably in this study.} therefore, might best be understood as the threshold separating synthetic media consumption from material impact.

\begin{figure}
  \centering
  \includegraphics[width=0.8\linewidth]{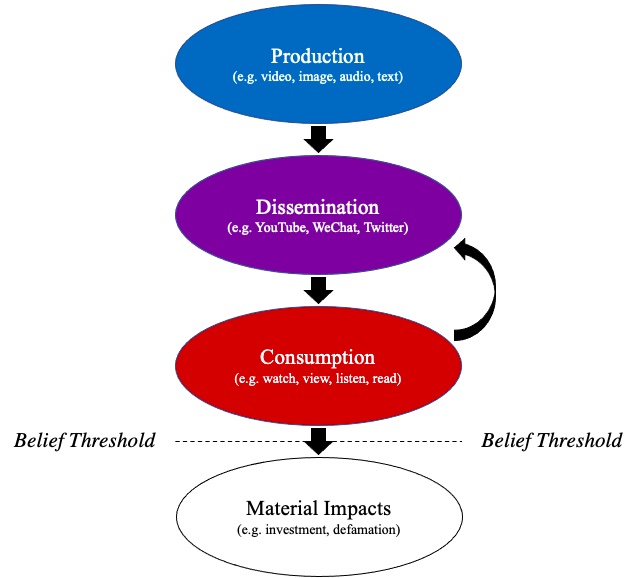} 
  \caption{The relationship between synthetic media, belief, and material impact.}
  \label{fig:example}
\end{figure}

Yet despite the established psychological importance of belief on actions (Locke, 1982), as well as its intuitive importance linking synthetic media and material impact, few studies have chosen to focus on this axis of synthetic media effects. Instead, most research tends to focus on Turing tests, examining the proportion of human subjects or algorithms that correctly distinguish genuine from synthetic content (Bray et al., 2022; Korshunov \& Marcel, 2020; Mai et al., 2023; Brown et al., 2020; Iacobucci et al., 2021; Guo et al., 2023). This type of test, while interesting, does not necessarily map onto the level of credibility a subject possesses about the media they are consuming. 

Furthermore, the few studies that do investigate belief have isolated themselves to the evaluation of \textit{political} content credibility (Hameleers et al., 2022; Kreps et al., 2022; Lee \& Shin, 2022). These studies ignore the many other information environments within which information gatherers might encounter synthetic media, despite initial evidence that synthetic textual media may possess influential strength that varies across subject domains -- such as medicine, psychology, and finance (Guo et al., 2023). 

For instance, one study involving a prominent text detection algorithm called “Grover” (Zellers et al., 2020) reported over 96\% classification accuracy for all news categories except one: financial news (Zellers, 2019). Grover’s performance on synthetic financial news was described as “a notable outlier,” with an accuracy of only 60.5\% (Zellers, 2019). However, the reason for this deviation was left empirically undefined.

Another paper compared the “helpfulness” \begin{CJK*}{UTF8}{gbsn}有用性\end{CJK*} rating of ChatGPT Q\&A responses to human responses across a variety of different subject domains (B. Guo, personal communication, 2023). Interestingly, synthetic responses in the finance category were rated at least 8\% more helpful than human responses in any other standalone subject. This trend held true for Q\&A comparisons in both Chinese and English (Guo et al., 2023). Furthermore, ChatGPT’s finance responses were the most difficult for “expert respondents” (familiar with the ChatGPT platform) to distinguish from human responses, when compared side-by-side. This trend also held true across language comparisons (Guo et al., 2023). 

\subsubsection{The Financial World’s Vulnerability to Synthetic Text}

The asymmetrical attention given to political synthetic media within preexisting literature aligns with the fact that synthetic use-cases overwhelmingly involve political subject matter (as seen in Section 2.3). This trend also corresponds with the relatively data-rich target profile of political figures, as well as the highly ``resource restrictive” nature of synthetic media production discussed in Section 2.6.1. Indeed, given the intensive time, capital, data, and computing requirements of “tailored” content generation (Helmus, 2022; Hwang, 2020), as well as the risk associated with criminal attribution (Hwang, 2020), one should not be surprised to see higher-quality (i.e. non-pornographic) synthetic media levied by and against resource-laden government actors. However, this characteristic of synthetic content generation also begs another question: where else should one expect synthetic media to emerge? 

While malicious uses of synthetic media within the financial realm are limited to ``only a handful of documented cases to date” (Bateman, 2020), several factors make it particularly vulnerable to synthetic media – especially text modalities. First, global financial markets are significantly influenced by ``investor sentiment,” (Baker \& Wurgler, 2007) which involves ``a belief about future cash flows and investment risks” (Baker \& Wurgler, 2007). Indeed, market sentiment has historically been attributed with multiple noteworthy deviations from rational financial models, to include the stock market crashes of 1929 and 1987, as well as the ``Dot.com” bubble burst in the 1990s (Baker \& Wurgler, 2007). This characteristic of the financial world is particularly concerning given that textual media inputs, such as financial news and social media commentary, are a major input to investor sentiment monitoring. Indeed, one of the world’s most ``heavily used and highly regarded professional investment system[s]” (Kenton, 2022) – known as the ``Bloomberg Terminal” – currently incorporates Twitter feeds within its platform for this very reason (“Bloomberg Launches a Twitter Feed Optimized for Trading | Press | Bloomberg LP,” 2018), and trading algorithms are often leveraged by fintech companies and high-frequency traders (Bateman, 2020).

Second, the risk tolerance available to actors within the financial domain mimics those of the political realm. Both commonly possess or can access resources sufficient enough to bypass the constraints of the synthetic media production stage, including expertise and capital. Additionally, both environments offer high-reward opportunities that represent a significant return on the intensive resource costs of synthetic content creation, as well as the criminal attribution risk that might accompany it. While this may translate to elections and power positioning in the political world, hundreds of billions of dollars in market movement may be at stake in the financial world. 

Finally, both political and financial information environments offer troves of public data available for training synthetic media algorithms. In the political world, public figures and debates are often available online, just as financial statements and updates to investors legally required by regulators expose publicly-listed corporations to synthetic media targeting (SEC.Gov | Financial Reporting Manual, n.d.; Estevez, 2022; Bateman, 2020). 

Several recent demonstrations of financial disinformation showcase the vulnerability of financial information environments to synthetic media. One particularly illustrative example of investor sentiment manipulation seeking \textit{upward} market movement occurred in 2021. The ``Wallstreet Bets" forum on Reddit, which was created to discuss financial trades online, made international headlines when a group of netizens swayed swaths of investors to drive up the stock price of used-video-game-company GameStop (Phillips \& Lorenz, 2021). Over the course of several days, GameStop became ``the most actively traded stock on Wall Street” (Phillips \& Lorenz, 2021), raising its market value by 1200\% to \$24 billion USD -- all while the general market declined (Phillips \& Lorenz, 2021). While this event may appear anecdotal in isolation, investor sentiment fluctuations spurred by the Wallstreet Bets forum has repeatedly influenced market movements since then, including a 280\% increase in the share price of software-company Blackberry, and an 840\% increase in entertainment-company AMC (Phillips \& Lorenz, 2021). 

A second example of investor sentiment manipulation targeted at \textit{downward} market movement is the wave of falsified social media accounts and messages that overtook Twitter in 2022, mimicking legitimate corporate entities. One account, impersonating pharmaceutical giant Eli Lilly, tweeted that it would begin providing insulin free of charge – one of Eli Lilly’s major products (Harwell, 2022). While the post was eventually removed by Twitter moderators, Eli Lilly’s stock price nonetheless fell by over 4\% the following day (Lee, 2022), amounting to ``billions” of dollars lost in market value (Harwell, 2022). Meanwhile, other companies that fell victim to disinformation posted from fake corporate accounts included Nintendo, Tesla, and PepsiCo (Mac et al., 2022; Alba, 2022). 

As seen above, the introduction of synthetic media to market manipulation efforts is a troubling prospect. As identified in the Carnegie Endowment for Peace’s report on ``Deepfakes and Synthetic Media in the Financial System,” attacks could broadly target entire markets (e.g. a ``flash crash”), or narrowly target specific brands (e.g. via ``botnets”) (Bateman, 2020). However, a ``lack of data” on historical attacks renders the magnitude of future threats in the financial arena ``difficult to judge” (Bateman, 2020).

\subsection{Hypotheses}

Given the aforementioned context, this study sought to more empirically examine the financial community’s vulnerability to synthetic media through an experiment on investor sentiment manipulation. Through the use of a survey, data was gathered on how much synthetic financial media shifted investor sentiment. Rather than merely asking investors directly, this experiment’s survey also gauged belief through a series of behavioral questions, including price change predictions and willingness to invest. Given the heightened importance of textual data toward investors and financial institutions, this experiment specifically focused on synthetic text. The experiment proceeded by examining several hypotheses:\newline 

\begin{itemize}
    \item \textbf{H1}: GPT-3.5 produces \textit{synthetic} financial disinformation that is more persuasive than \textit{genuine} financial information.
\end{itemize}

Two factors drove the reasoning behind this hypothesis. First, financial reports are more standardized than creative forms of text, given reporting requirements to investors (SEC.Gov | Financial Reporting Manual, n.d.). This might reasonably allow LLMs to more easily modify document content without risking document structure. Second, financial text relies heavily on high-context numerical figures, which are difficult to observe and easy to manipulate. For instance, unless an investor were standing directly in a manufacturing facility, it would be difficult to assess an inventory turnover metric. By comparison, falsified world events or sports news might be easier for readers to assess from afar. 

One real-world demonstration of this principle occurred during Microsoft’s presentation of its OpenAI-powered search engine: “Bing AI” (Mehdi, 2023; Warren, 2023). During the demo, one of Microsoft’s executives showcased Bing AI’s capabilities by generating summarized financial reports. Only after the presentation did online viewers realize that the search engine had actually offered multiple severely inaccurate metrics as fact (Warren, 2023). For instance, Bing AI indicated that clothing-company Gap held an operating margin of 5.9\% in Q3 2022, despite the fact that this figure did not appear anywhere in Gap’s financial statements (Warren, 2023). It also inflated earnings per share by over 60\%, doubled Gap's cash reserves, and miscalculated inventory by over 30\% (Leswing, 2023). Unless a viewer was intimately aware with a company's earnings, or the falsification was orders of magnitude removed from a realistic figure, verification of such financial claims would likely be difficult to assess in real-time.

For clarity, this study’s primary hypothesis did not seek to draw comparisons between synthetically produced financial media and other forms of synthetically produced media -- such as the world news article contained in "All the News That's Fit to Fabricate" (Kreps et al., 2022). However, this question is thought-provoking and is thus left as an exploratory sub-hypothesis:

\begin{itemize}
  \item \textbf{H2}: GPT-3.5 produces synthetic \textit{financial} disinformation that has more persuasive power than synthetic \textit{world news} disinformation.
\end{itemize}

Lastly, if the assumptions in H1 are true, then it is reasonable to believe that GPT-3.5 might produce more persuasive financial content for niche industries than non-niche industries. For example, the average investor might be more easily misled about semiconductor component manufacturing than a produce retailer she visits on a weekly basis. Although this question is not evaluated with statistical rigor due to the time constraints of the study, it is left here for consideration in future research, and is touched upon when possible:

\begin{itemize}
  \item \textbf{H3}: Synthetic financial disinformation about \textit{niche industries} is more persuasive than that of \textit{non-niche industries}.

\end{itemize}

\subsection{Methodology}

In order to begin empirically investigating the effects of synthetic media on global financial communities, this study piloted an exploratory field survey in China. Because many foreign media applications are officially banned in China, this study leveraged the survey mini-app, Wenjuanxing \begin{CJK*}{UTF8}{gbsn}问卷星\end{CJK*}, via China’s largest social media application: WeChat \begin{CJK*}{UTF8}{gbsn}微信\end{CJK*}. Participants were sampled from China’s financial services community and were each paid 50 RMB (approximately \$7.30 USD) to randomly complete a survey containing one of three treatment texts. The control group received a genuine, Q2 2018 earnings call transcript from beverage manufacturer The Coca-Cola Company (henceforth “Coca-Cola”). The second treatment group received a ChatGPT-generated Coca-Cola earnings call transcript for the same fiscal period. Finally, the third treatment group received a ChatGPT-generated earnings call transcript of PC manufacturer HP Inc. (henceforth “HP”). This was done to explore inter-industry differences between synthetic financial texts (sub-hypothesis H3).

Participants were informed that the survey was targeted at assessing how different kinds of financial news influence readers, were instructed not to rely on outside materials for question-answering, and were asked not to change responses once they had been completed. Additionally, the terms “real,” “fake,” and “ChatGPT” were deliberately not mentioned within the survey in order to avoid biasing participants.

\subsubsection{Participants}

The decision to exclusively sample respondents within the financial services community was intended to maximize the likelihood of response completion on topical survey content (i.e. financial markets), and to approximate a population of experienced investors as closely as possible. For context, only 7\% of Chinese citizens owned stock in 2019 compared to 53\% of Americans (Johnston, 2022). The Chinese government also regulates investment in companies overseas (Zhou \& Leng, 2021). Although this decision scientifically limited the generalizability of this survey’s results, sampling a financially literate sub-population was determined as both cost-efficient (given limited available funding), and a best alternative for accurately representing global investors (given the survey team's physical location in China). Out of the 165 participants originally surveyed, 19 incorrectly answered an attention-check question, yielding a final sample size of 146 respondents.

\begin{figure}
  \centering
  \includegraphics[width=0.8\linewidth]{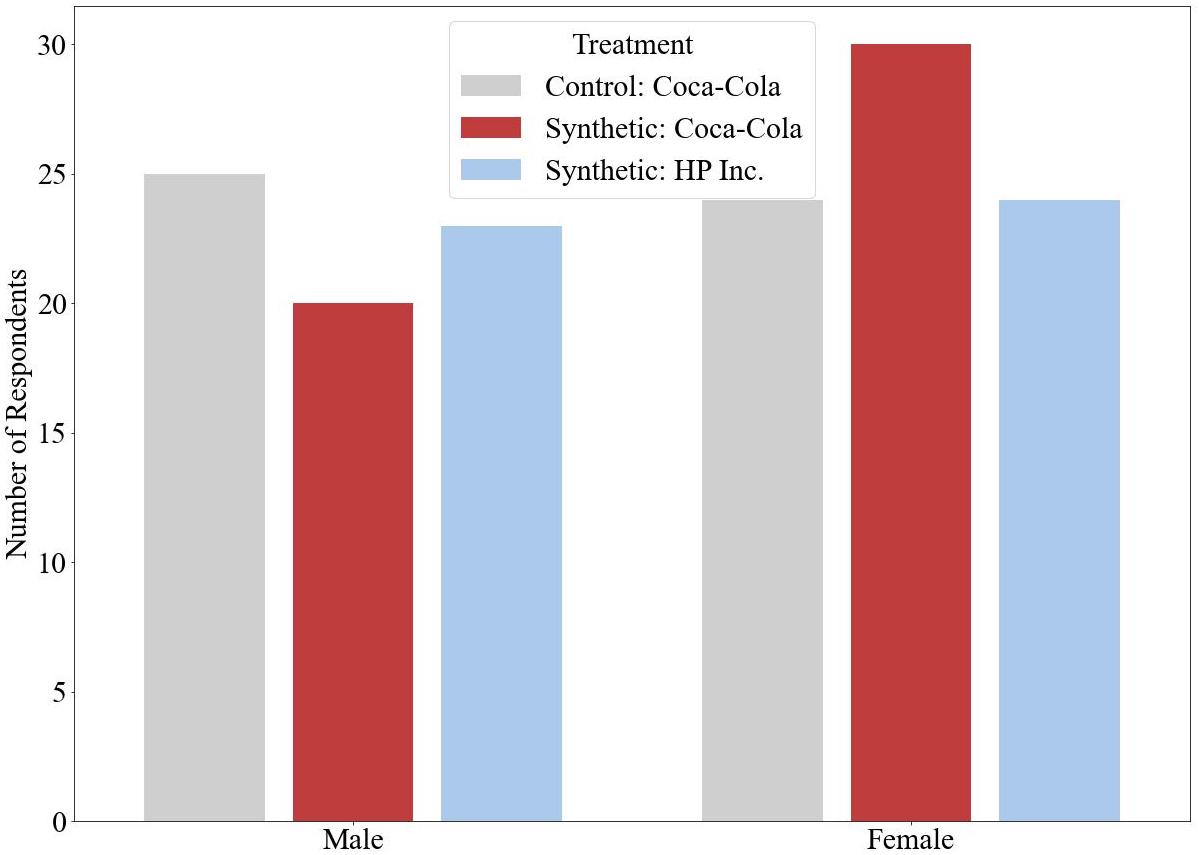} 
  \caption{Gender distribution across treatment groups.}
  \label{fig:gen}
\end{figure}

\begin{figure}
  \centering
  \includegraphics[width=0.8\linewidth]{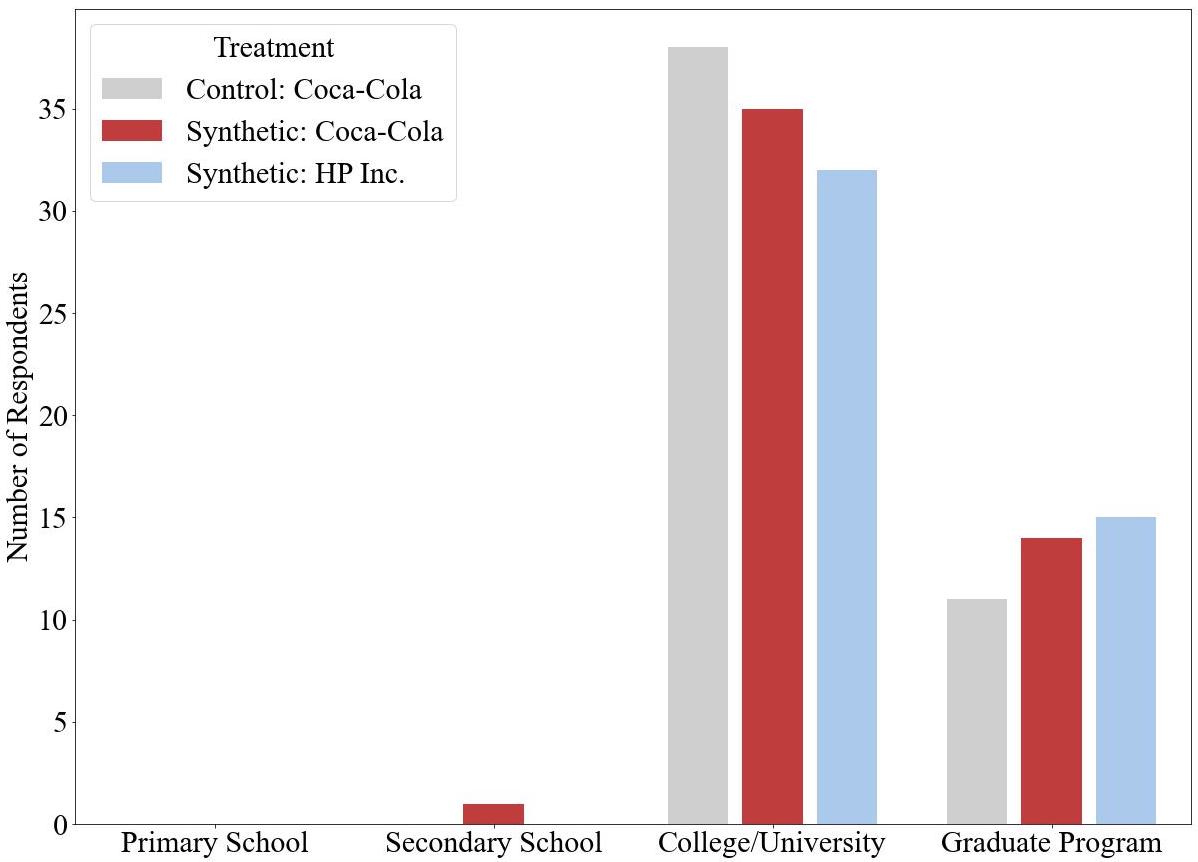} 
  \caption{Education level distribution across treatment groups.}
  \label{fig:ed}
\end{figure}

\begin{figure}
  \centering
  \includegraphics[width=0.8\linewidth]{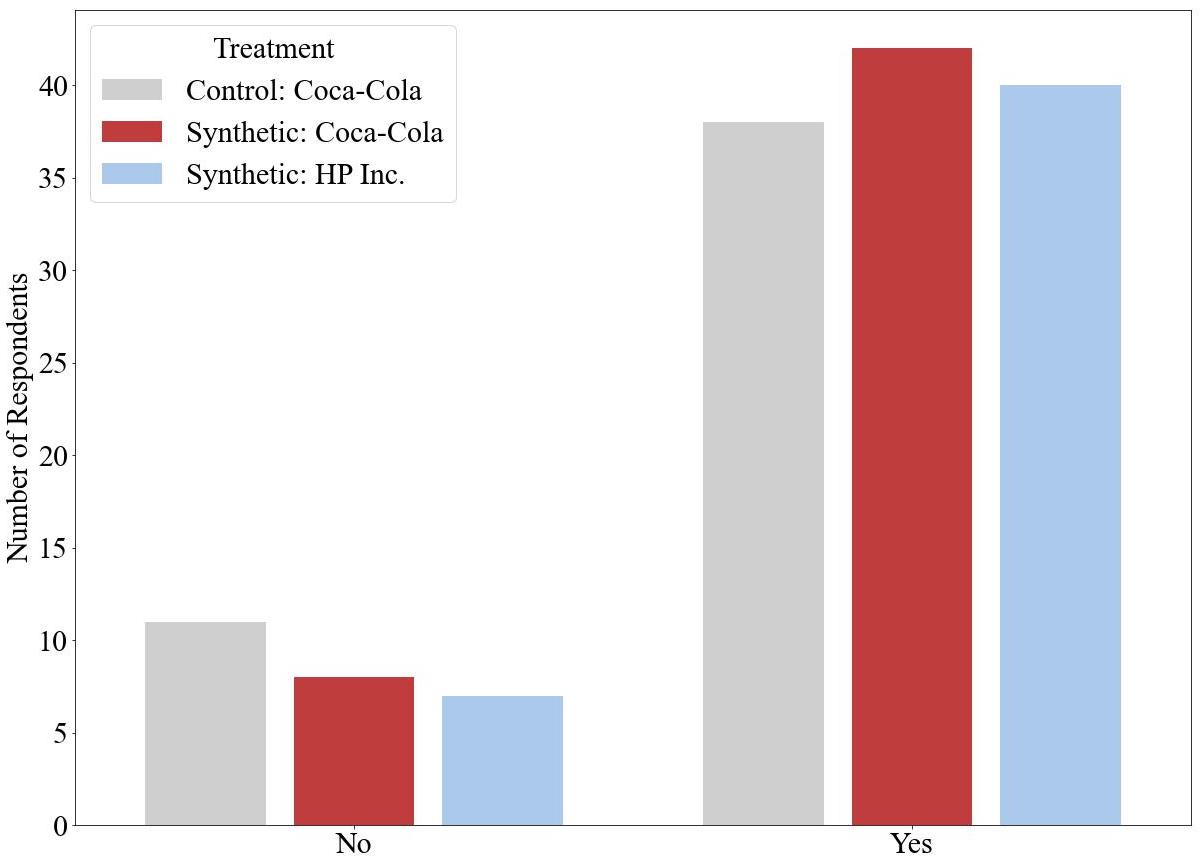} 
  \caption{Financial services experience across treatment groups.}
  \label{fig:finser}
\end{figure}

\begin{figure}
  \centering
  \includegraphics[width=0.8\linewidth]{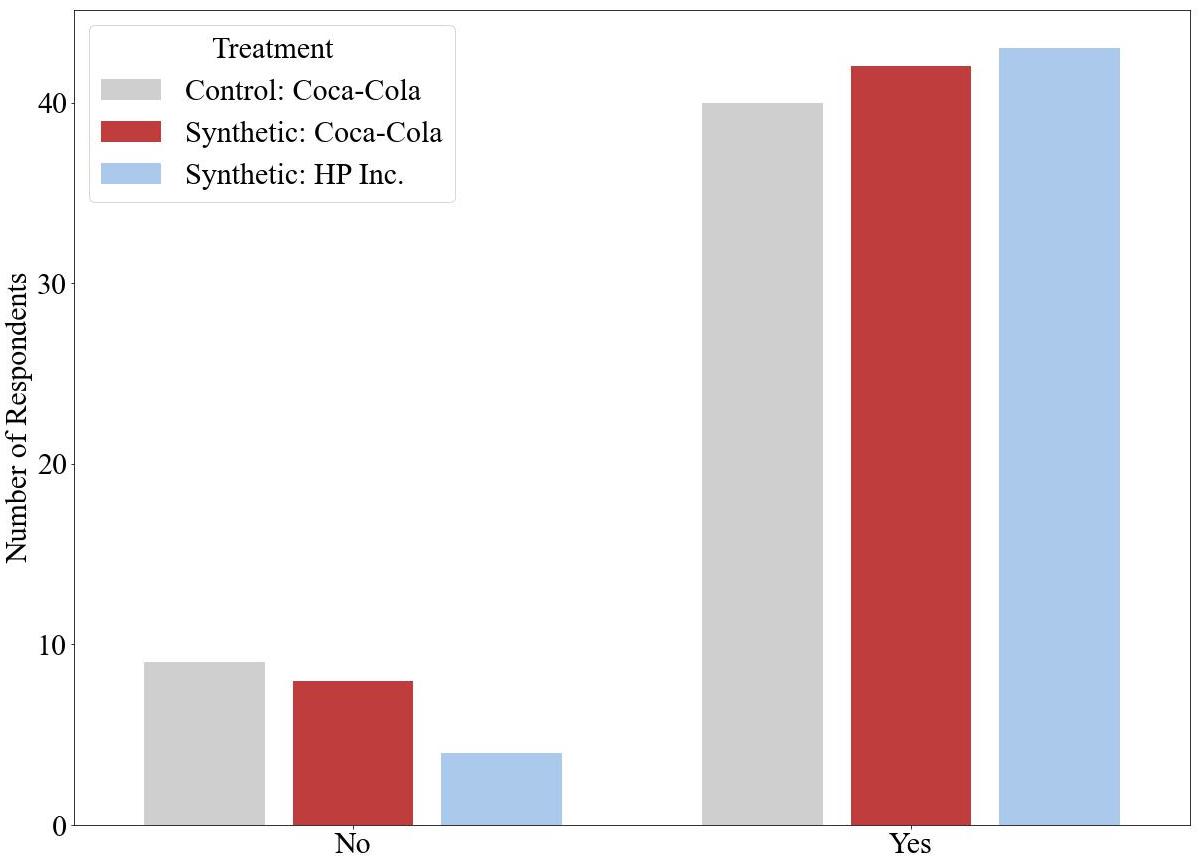} 
  \caption{Investment experience across treatment groups.}
  \label{fig:invest}
\end{figure}

\subsubsection{General Design}

All three survey versions possessed the same structure. Each contained two pre-treatment questions, a treatment text, and 17 post-treatment questions. Treatment texts were also presented in the same format: an abbreviated company ``earnings call” transcript.

\subsubsection{Earnings Call Transcripts}

For clarification, earnings calls are press releases conducted by public companies on  a quarterly basis. During these calls, a company’s top executives share quarterly progress with investors directly and conclude by hosting a Q\&A session with financial analysts about the company’s future economic outlook. While executives are expected to report factual information in accordance with the U.S. Securities and Exchange Commission’s (SEC) reporting requirements, earnings calls often ``cast the previous quarter in the best light possible” (Estevez, 2022). Analysts track these calls closely to gauge whether a given company’s quarterly expectations were met. Consequently, ``there is usually a direct and immediate response to [a company's] stock price” following an earnings call’s conclusion (Estevez, 2022).

The control treatment in this study utilized a publicly-available transcription of Coca-Cola's Q2 2018 earnings call via The Motley Fool's website (Coca-Cola Company (The) (KO) Q2 2018 Earnings Conference Call Transcript, 2018).

Several design decisions are worth explaining in more detail. For instance, four factors drove the decision to use an earnings call transcript from 2018. First, prompting ChatGPT for content after 2021 resulted in errors, due to the model’s training data “cutoff” (Why doesn't ChatGPT know about X? | OpenAI Help Center, n.d.). Therefore, in order to restrict variation between control and experimental treatments, genuine earnings call transcripts had to be sampled prior to 2021. Second, the use of a 2018 earnings call facilitated the possibility of results comparison against earlier language models (trained on pre-2019 data) in future research studies. Third, utilizing a non-recent earnings call intuitively reduced the possibility that respondents were able to rely on memory to recall company performance, better isolating treatment effects on respondent behavior. Finally, using a pre-2019 earnings call removed confounding associations with COVID-19's influence on equity markets. In the spirit of this consideration, a second quarter earnings call was also chosen to avoid the possibility of volatile investor sentiment at the beginning and end of the 2018 fiscal year.  

The use of Coca-Cola and HP as treatment companies was deliberate as well. Choosing multinational corporations with a brand presence in China seemed to mitigate the general lack of Chinese investment activity in international securities. Thus, company names were randomly sampled from “The S\&P Global 100 Index” -- a list including the top multinational enterprises in the world (S\&P Global 100, n.d.; S\&P Global 100 Constituents, n.d.). Coca-Cola and HP both exist within this global ranking. From there, candidate companies were manually culled using several criteria. First, the candidate in question had to have a 2018 Q2 or Q3 earnings call transcript posted on The Motley Fool’s website. Second, the candidate had to have a presence in China’s domestic market, in order to cater to respondents’ visibility and awareness of the company. As of 2013, Coca-Cola owned 16\% of China’s beverage market share (Coca-Cola Says to Invest over \$4 Billion in China in 2015-2017 | Reuters, 2013), and as of 2023, HP has operated in China for 40 years (Ma, 2023). Finally, the transcript had to be free of transcription errors by The Motley Fool, such as the indication of “inaudible” phrases, in order to focus survey responses on transcript content rather than syntax errors. 

For both genuine and fake earnings call transcripts used in this study, only the first 550 Chinese characters (after translation) were considered usable, with a buffer of 50 characters granted to prevent unnatural, mid-paragraph cutoffs in the treatment text. This standardization was chosen because the free version of ChatGPT limits output to about 500 words, as well as to avoid excessively biasing survey respondents towards text length. Earnings call cruft – such as call operator comments, call handovers, etc. – was also removed in accordance with the methodological notion of ``upper-bound credibility” adopted from the study, ``All the News That’s Fit to Fabricate” (Kreps et al., 2022). ``Upper-bound credibility” conditions reasonably assume that disinformation campaign actors will ``cherry pick” autonomously generated content that maximizes the likelihood it will fool its audience (Kreps et al., 2022). Overall, seven words were excluded from the beginning of the genuine Coca-Cola transcript (``James Quincy: Thanks, Tim. Good morning, everyone.” \begin{CJK*}{UTF8}{gbsn}詹姆斯昆西\end{CJK*}: \begin{CJK*}{UTF8}{gbsn}谢谢\end{CJK*}, Tim. \begin{CJK*}{UTF8}{gbsn}大家早上好\end{CJK*}), and an incompatible translation for ``mid-single-digit" was converted to ``4\%-6\%."\footnote{As discussed in later sections, metrics may have played an important role in influencing survey respondents. However, under these assumptions, this alteration to the control treatment text should have \textit{weakened} the experiment's results, if anything. Therefore, it is ignored in the discussion of experimental ``Limitations."} For the synthetic Coca-Cola transcript, one awkward phrase was removed (“...\begin{CJK*}{UTF8}{gbsn}和某些市场单位箱体积下降\end{CJK*}.”), and two awkward character-phrases were adjusted (\begin{CJK*}{UTF8}{gbsn}在\end{CJK*} changed to \begin{CJK*}{UTF8}{gbsn}可以\end{CJK*}, and \begin{CJK*}{UTF8}{gbsn}福利\end{CJK*} changed to \begin{CJK*}{UTF8}{gbsn}生活方式\end{CJK*}), based on translator recommendations. CEOs typically speak first during earnings calls, so all three abbreviated treatment texts exclusively contained CEO commentary. Ultimately, treatment texts were 547 words long, on average.\footnote{The number of Chinese characters is not perfectly equivalent to word count (given that some Chinese phrases consist of two or more characters). However, it serves as rough approximation, and was thus treated as equivalent.}

\subsubsection{Survey Questions}

Emulating "All the News That's Fit to Fabricate," survey respondents were explicitly asked to rate the ``credibility" (Q5) of a given treatment text (Kreps et al., 2022). However, several \textit{behavioral} survey questions were also included to more comprehensively measure the influence of synthetic financial text on investors. These included short-term economic outlook of the treatment company before and after treatment (PQ1, Q3),\footnote{``Q3," here, stands for ``Question 3," and should not be confused with similar abbreviations for fiscal quarters. ``PQ1" stands for ``Pre-Treatment Question 1."} prediction in share price change (Q1), and willingness to invest (Q2).

Because treatment groups contained approximately 50 observations each, a ``student t-distribution" was used to compare sample statistics. Statistical significance for survey responses was thus determined using independent, two-sample t-tests.\footnote{Although ``Change in Short-Term Economic Outlook" score was calculated within treatment samples, treatment groups were treated as independent samples. Therefore, the difference in means between treatment scores was also calculated using an independent, two-sample t-test.} Additionally, Welch’s t-tests were used to account for unequal variance across groups. 

Although data was gathered on an HP treatment text to begin exploring sub-hypothesis H3, the primary focus of this study was to gauge differences between the control and treatment Coca-Cola texts. Therefore, statistical analysis was constrained to these two treatment groups (with exception to inter-synthetic treatment comparisons). The null hypotheses for each of the research questions posed in this study amount to the absence of meaningful difference in persuasive power between control and treatment texts.

\subsubsection{Procedure}

In order to cater to Chinese respondents, survey questions for each survey version were translated from English into Chinese via ChatGPT and human editing (OpenAI, 2022). However, in order to avoid introducing bias between genuine and synthetic text, treatment texts were translated via different methods. The genuine earnings call transcript chosen for this study was translated from English into simplified Chinese via two human translators – both possessing background knowledge in finance. 

Synthetic transcripts, on the other hand, were generated via the March 14, 2023, version of ChatGPT (OpenAI, 2022). Due to ChatGPT’s inability to consistently generate Chinese text directly, as well as OpenAI’s disproportionate use of English-language data to train its language models (Brown et al., 2020), synthetic treatment texts were initially generated in English before translation back into Chinese – all via ChatGPT (OpenAI, 2022).\footnote{When asked to generate Chinese text \textit{directly}, ChatGPT haphazardly produced an automatic error about being an “English language model” (OpenAI, 2022). Therefore, ChatGPT was first prompted to generate synthetic treatment texts in English before translating them into Chinese.} The quality of the synthetic Coca-Cola transcript was also secondarily vetted by translators, in order to ensure conditions of upper-bound credibility were upheld.

Coca-Cola's synthetic transcript was generated via the following prompt: ``Generate the first 550 words of a fictional Q2 2018 earnings call transcript for The Coca-Cola Company with a negative economic outlook.” HP’s synthetic transcript was generated by simply replacing ``The Coca-Cola Company” with ``HP Inc.," as well as an explicit instruction to ``restrict output to comments by the CEO" to match the Coca-Cola output format. The prompt modifier ``with a negative economic outlook” was deliberately chosen to juxtapose the tendency of genuine earnings calls to frame company economic outlooks in “the best light possible” (Estevez, 2022). This enabled a clearer examination of changes in respondent perception of their treatment company before and after treatment. Translations into simplified Chinese were generated by instructing ChatGPT to “Translate the following into Chinese:”, followed by pasting the full, synthetic transcript in its English version. 

\section{Results}

\subsection{Change in Short-Term Economic Outlook (PQ1, Q3)}

In order to account for biases, survey participants were asked two pre-treatment questions about how they perceived the short-term (PQ1) and long-term (PQ2) economic outlooks of their treatment company. Response options mimicked the four-option Likert scale utilized in “All the News That’s Fit to Fabricate” (Kreps et al., 2022): “Very unfavorable” \begin{CJK*}{UTF8}{gbsn}非常不利\end{CJK*}, "Somewhat unfavorable” \begin{CJK*}{UTF8}{gbsn}有点不利\end{CJK*}, "Somewhat favorable \begin{CJK*}{UTF8}{gbsn}有点有利\end{CJK*}, and “Very favorable” \begin{CJK*}{UTF8}{gbsn}非常有利\end{CJK*}.

Following treatment, participants were asked to assess the short-term economic outlook of the treatment company again. Participants in the control group considered Coca-Cola's new economic outlook to be “somewhat favorable” on average (mean: 3.163, SD: $\pm$ 0.681), while participants in the synthetic Coca-Cola group (mean: 2.260, SD: $\pm$ 0.626) generally held less favorable views following treatment.

\begin{figure}
  \centering
  \includegraphics[width=0.8\linewidth]{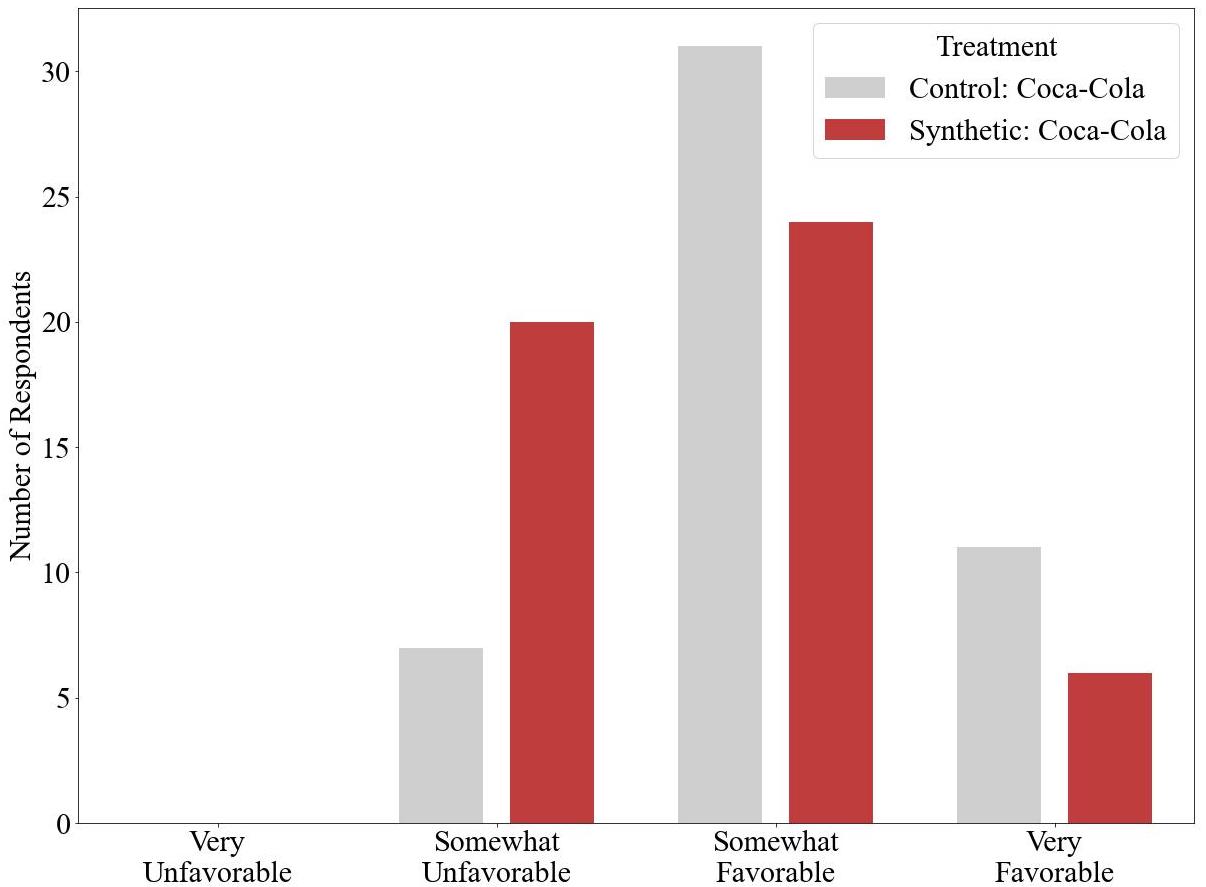} 
  \caption{Old short-term economic outlook (PQ1).}
  \label{fig:finser}
\end{figure}

\begin{figure}
  \centering
  \includegraphics[width=0.8\linewidth]{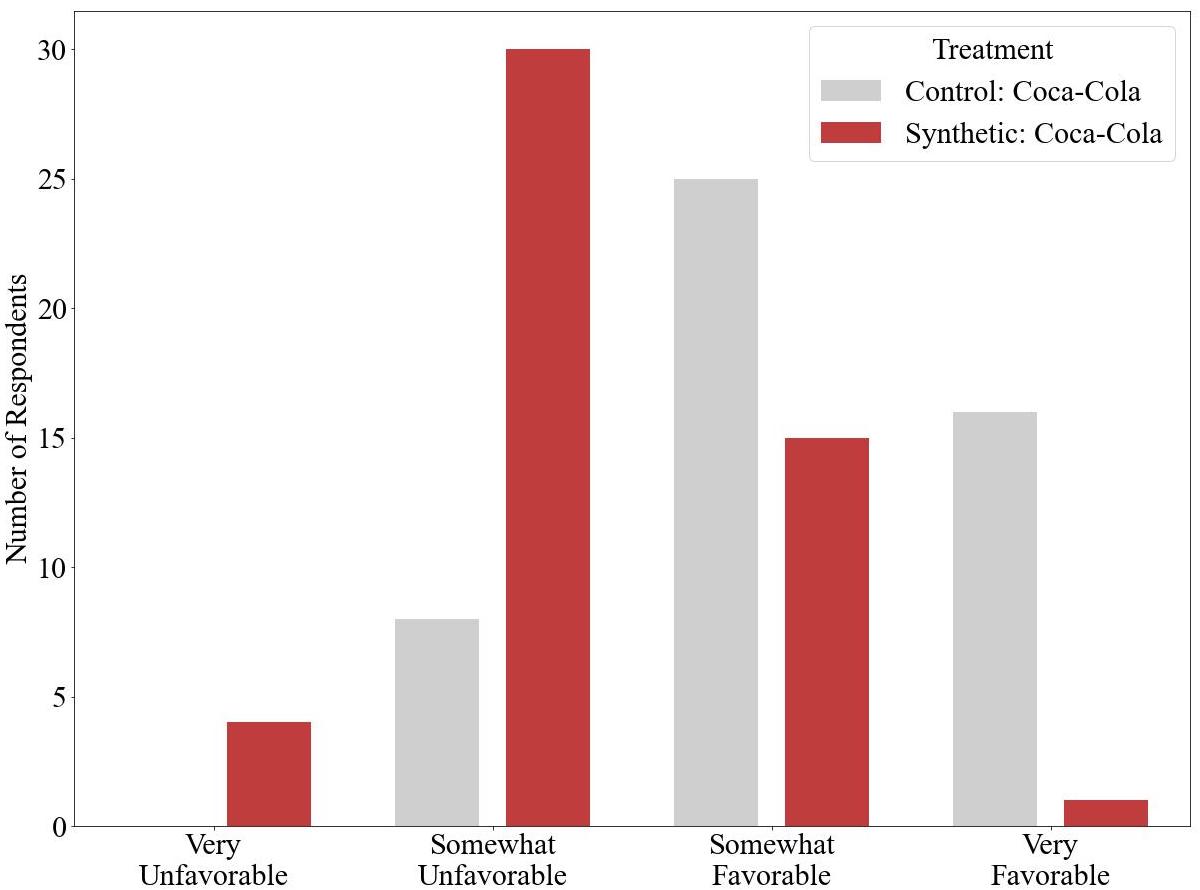} 
  \caption{New short-term economic outlook (Q3).}
  \label{fig:invest}
\end{figure}

The average difference in (raw) outlook scores after treatment amounted to over a half a step down the Likert scale (difference in means: -0.542, p-value: 0.000{*}{*}{*}), indicating that exposure to a (negative) synthetic earnings call significantly damaged participants’ perspective of Coca-Cola's financial outlook in the near-term, compared to the control group. Of course, this might be expected given the negative prompt modifier provided to ChatGPT prior to synthetic transcript generation.

\begin{table}
  \centering
  \begin{threeparttable}[c]
    \caption{Outlook Difference Between Treatments (Raw Score)}
    \label{tab:three-part-table-q3}
    \begin{tabular}{ c c c c }
      \toprule
      Treatment & Mean & Diff. in Means & p-value \\
      \midrule
      Control & 0.082 & -- & -- \\
      \hline
      Synthetic & -0.460 & -0.542 & 0.000{*}{*}{*} \\
      \bottomrule
    \end{tabular}
    \begin{tablenotes}
      \item [a] {*}p < 0.05, {*}{*}p < 0.01, {*}{*}{*}p < 0.001.
    \end{tablenotes}
  \end{threeparttable}
\end{table}

In order to evaluate the influential power of synthetic treatment independently of positive or negative valence, two additional tests were undertaken on the data. First, a t-test was taken using the \textit{absolute value} of change in outlook scores across treatment groups. Second, an equivalent t-test was taken, but on a \textit{binary} ``conversion” variable. This latter method was inspired by the methodology in “All the News That’s Fit to Fabricate,” in which four-option Likert scale responses were condensed into two binary categories (Kreps et al., 2022).\footnote{“Non-credible” and “Credible” (Kreps et al., 2022).} In this case, “Very unfavorable” and “Somewhat unfavorable” responses where grouped together in a singular “Unfavorable” category, while “Somewhat favorable” and “Very favorable” were coded as “Favorable” responses. A respondent’s change from one category to another category after treatment yielded a score of 1, while no change corresponded to a score of 0. This engineered ``conversion" feature selectively rewarded extreme changes in perspective, rather than incremental shifts. Conveniently, this numerical mean also corresponded to the percentage of responses within each treatment group, improving interpretability.

While the first method did not show differences across treatments, the binary conversion metric yielded a 17.6\% increase in (average) outlook conversions across treatments (difference in means: 0.176, p-value: 0.049{*}). These results were robust against correlation with long-term economic outlook, and lend support to the primary hypothesis (H1) that ChatGPT-generated financial texts may have greater influence over investors than genuine financial texts do. Of course, these findings are currently constrained to a sample of Chinese investors, the use of GPT-3.5, and earnings call text involving The Coca-Cola Company. However, the possibility that synthetic text holds powerful sway over investors carries sobering implications for the global financial community. 

\begin{table}
  \centering
  \begin{threeparttable}[c]
    \caption{Outlook Difference Between Treatments (Binary Score)}
    \label{tab:three-part-table-bin}
    \begin{tabular}{ c c c c }
      \toprule
      Treatment & Mean & Diff. in Means & p-value\\
      \midrule
      Control & 0.184 & -- & -- \\
      \hline
      Synthetic & 0.360 & 0.176 & 0.049{*} \\
      \bottomrule
    \end{tabular}
    \begin{tablenotes}
      \item [a] {*}p < 0.05, {*}{*}p < 0.01, {*}{*}{*}p < 0.001.
    \end{tablenotes}
  \end{threeparttable}
\end{table}

\subsection{Share Price Change Prediction (Q1)}

The first post-treatment question asked survey participants how they anticipated treatment company share price to change (Q1), “based on [their] impression” of the treatment text. In order to set a clear baseline, the pre-treatment share price of the candidate company was hypothetically (and arbitrarily) established at \$100 USD within the question prompt. Participants could choose between five share price predictions: a decrease of more than \$2 USD, a decrease of less than \$1 USD, no change in price, an increase of less than \$1 USD, and an increase of more than \$2 USD. 

\begin{figure}
  \centering
  \includegraphics[width=0.8\linewidth, height=0.6\linewidth]{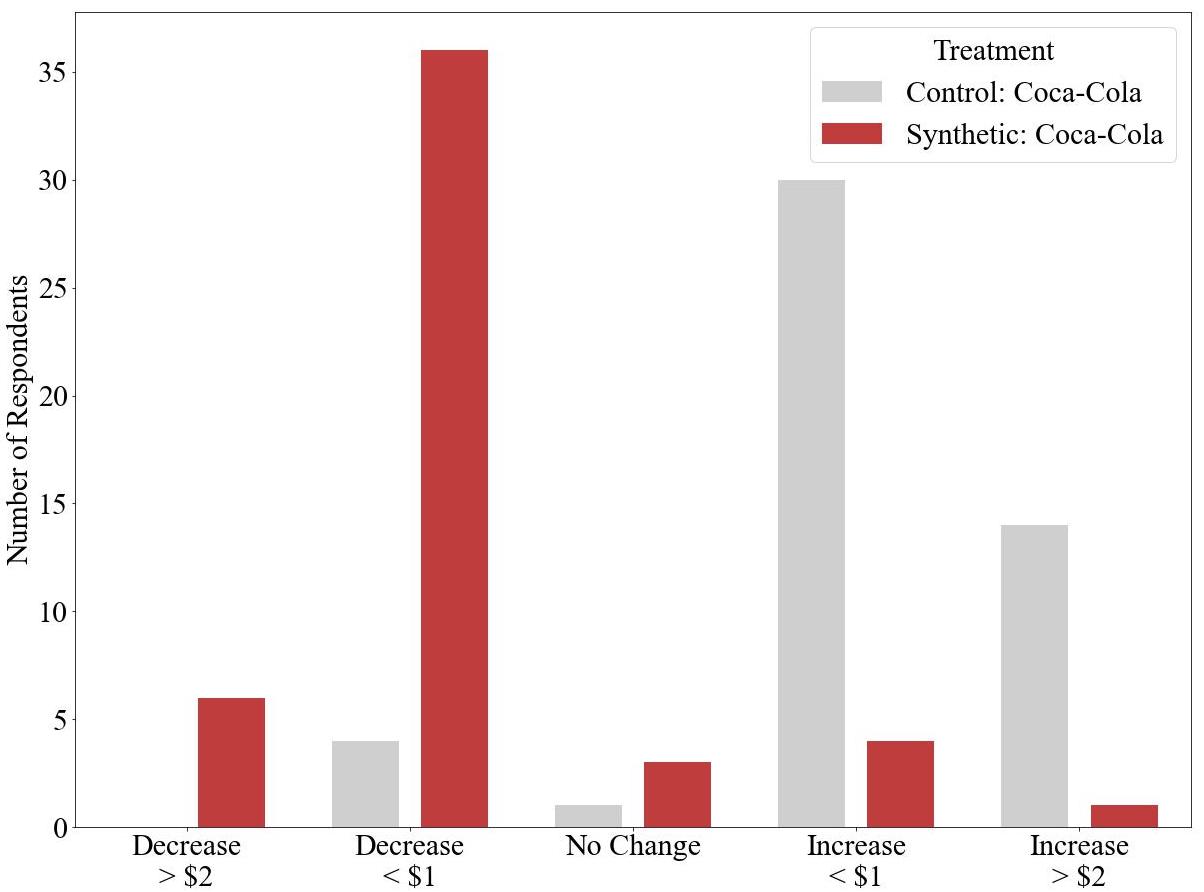} 
  \caption{Share price change prediction between treatments (Q1).}
  \label{fig:price}
\end{figure}

Respondents within the control group predicted a share price increase of less than \$1 USD on average (mean: 4.102, SD: $\pm$ 0.789), while the synthetic treatment group predicted a decrease of less than \$1 USD on average (mean: 2.160, SD: $\pm$ 0.809). These results produced a clear difference in means across treatment groups, with the experimental group demonstrating a decrease of almost 2 entire Likert scale levels (difference in means: -1.942, p-value: 8.100e-21{*}{*}{*}). These findings further solidified the results of Q3, regarding change in short-term economic outlook. The use of price prediction also forms a compelling starting point for quantifying impact (in dollar terms) within future research.

\begin{table}
  \centering
  \begin{threeparttable}[c]
    \caption{Share Price Prediction Difference Between Treatments (Raw Score)}
    \label{tab:three-part-table-price}
    \begin{tabular}{ c c c c }
      \toprule
      Treatment & Mean & Diff. in Means & p-value\\
      \midrule
      Control & 4.102 & -- & --\\
      \hline
      Synthetic & 2.160 & -1.942 & $8.1e^{-21}$ {*}{*}{*} \\
      \bottomrule
    \end{tabular}
    \begin{tablenotes}
      \item [a] {*}p < 0.05, {*}{*}p < 0.01, {*}{*}{*}p < 0.001.
    \end{tablenotes}
  \end{threeparttable}
\end{table}

\subsection{Investment Willingness (Q2)}

The second post-treatment question asked survey participants to place themselves in an imaginary scenario in which they received a \$1000 USD gift (Q2). The participants were then asked to choose among five options: either keep the \$1000 USD, or use portions of the gifted money to invest in the treatment company’s stock, broken down along \$250 USD increments (keeping the remainder as cash).

\begin{figure}
  \centering
  \includegraphics[width=0.8\linewidth, height=0.6\linewidth]{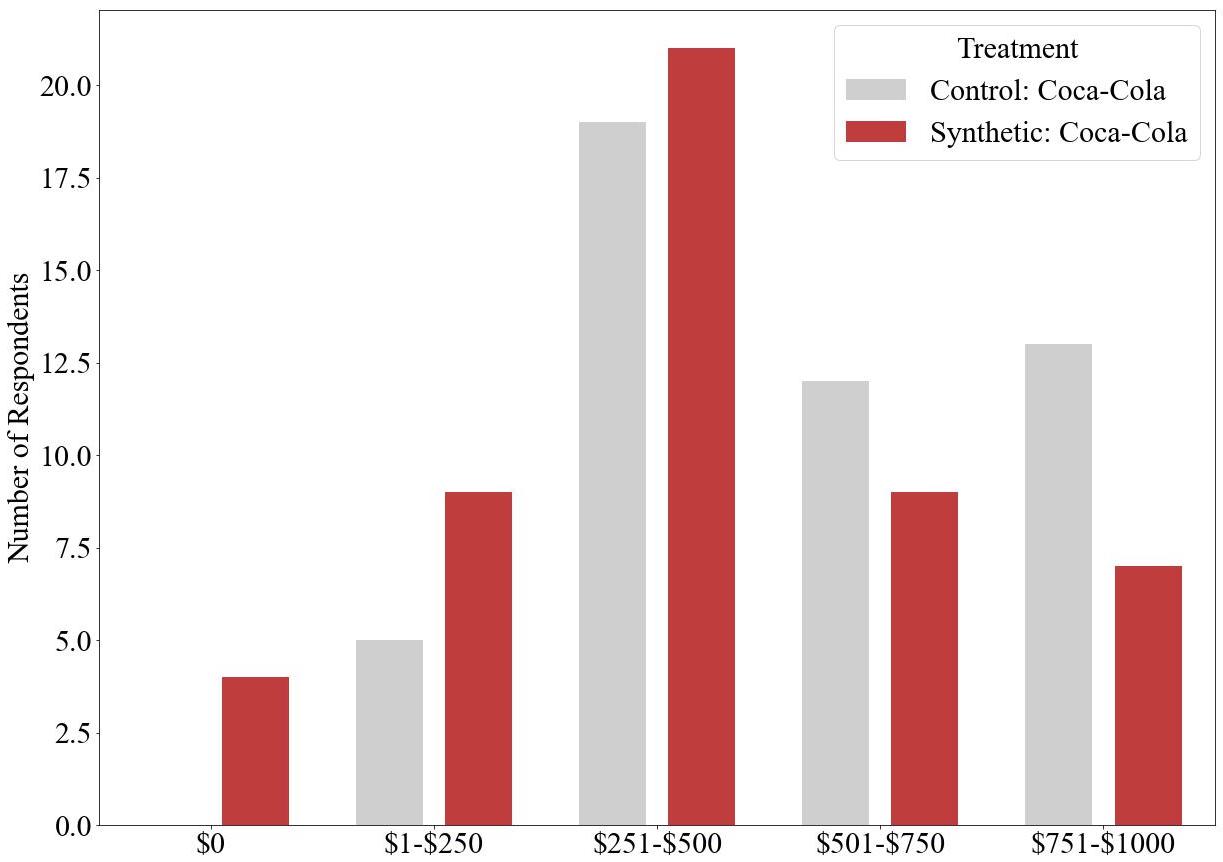} 
  \caption{Investment willingness between treatments (Q2).}
  \label{fig:investment}
\end{figure}

The synthetic treatment group (mean: 3.120, SD: $\pm$ 1.107) demonstrated diminished willingness to invest in their treatment company than the control group did, on average (mean: 3.673, SD: $\pm$ 0.977). Notably, all of the control group’s respondents chose to invest some dollar amount in Coca-Cola, while four synthetic group respondents chose to invest \$0 USD. Exposure to the synthetic treatment (on average) reduced investment willingness in Coca-Cola by over half a Likert level, compared to the control treatment (difference in means: -0.553, p-value: 0.010{*}). In other words, these findings support the claim that exposure to a synthetic earnings call reduced a participant’s investment willingness by approximately \$125, on average.

\begin{table}
  \centering
  \begin{threeparttable}[c]
    \caption{Investment Willingness Between Treatments (Raw Score)}
    \label{tab:three-part-table-invest}
    \begin{tabular}{ c c c c }
      \toprule
      Treatment & Mean & Diff. in Means & p-value \\
      \midrule
      Control & 3.673 & -- & -- \\
      \hline
      Synthetic & 3.120 & -0.553 & 0.010{*}\\
      \bottomrule
    \end{tabular}
    \begin{tablenotes}
      \item [a] {*}p < 0.05, {*}{*}p < 0.01, {*}{*}{*}p < 0.001.
    \end{tablenotes}
  \end{threeparttable}
\end{table}

Of note, survey participants also demonstrated, on average, a lower willingness to invest in HP than Coca-Cola after exposure to synthetic treatment (difference in means: -0.609, p-value: 0.009{*}{*}). These results suggest there is preliminary evidence for differences across financial texts by industry.

\begin{table}
  \centering
  \begin{threeparttable}[c]
    \caption{Investment Willingness Between Industries (Raw Score)}
    \label{tab:three-part-table-inter}
    \begin{tabular}{ c c c c }
      \toprule
      Treatment & Mean & Diff. in Means & p-value\\
      \midrule
      Synthetic: KO & 3.120 & -- & -- \\
      \hline
      Synthetic: HP & 2.511 & -0.609 & 0.009{*}{*} \\
      \bottomrule
    \end{tabular}
    \begin{tablenotes}
      \item [a] {*}p < 0.05, {*}{*}p < 0.01, {*}{*}{*}p < 0.001.
      \item [b] The "Synthetic: Coca-Cola" treatment served as the baseline control for this difference in means comparison.
    \end{tablenotes}
  \end{threeparttable}
\end{table}

However, the reasons behind this observed effect are uncertain and statistically challenging to verify within the scope of this study. Examining the relationship between pre and post-treatment variables indicates a potential confounding factor: long-term economic bias. Indeed, long-term economic outlook had three times more positive of a correlative effect on investment willingness towards HP (0.34) than towards Coca-Cola (0.10). 

A less favorable long-term outlook coupled by a lower investment willingness towards HP may reveal a  “fundamentalist” investment approach within the sampling group -- an investment philosophy which espouses a preference for holding long-term investments based on the durable value of a company rather than trading based on short-term changes in stock price (Zucchi, 2021). Given that HP Inc. is a technology company, it is possible that these results represent the caution of experienced investors towards higher volatility technology stocks. Confirming this relationship merits further research and a more robust sample size of investors.

\subsection{Credibility Assessment of Treatment Text (Q5)}

Leveraging a four-option Likert scale similar to the one found in “All the News That’s Fit to Fabricate" (Kreps et al., 2022), the fifth post-treatment question asked survey participants to directly rate the treatment text as either “Completely unbelievable" \begin{CJK*}{UTF8}{gbsn}完全不可信\end{CJK*}, “Somewhat unbelievable” \begin{CJK*}{UTF8}{gbsn}不太可信\end{CJK*}, “Somewhat believable” \begin{CJK*}{UTF8}{gbsn}有些可信\end{CJK*}, and “Very believable” \begin{CJK*}{UTF8}{gbsn}非常可信\end{CJK*}. Although Kreps et al.’s study does acknowledge the possible dissection of credibility into multiple dimensions, it also confirms the logical decomposition of “credibility” into “believability” (Kreps et al., 2022). In order to simplify analysis, this study asked participants to directly assess credibility as a standalone characteristic. Dividing the Likert scale into an even number of response options had the additional benefit of encouraging survey respondents onto either side of the response spectrum, rather than allowing them to occupy a neutral middle ground.

Two different methods were used for aggregating the results of this question. The first method merely examined the difference in raw credibility scores of the control group (mean: 3.265, SD: $\pm$ 0.442) and the synthetic treatment group (mean: 3.220, SD: $\pm$ 0.502). However, the results did not statistically support rejection of the null hypothesis that genuine and synthetic earnings call transcripts share comparable degrees of credibility. 

\begin{figure}
  \centering
  \includegraphics[width=0.8\linewidth, height=0.6\linewidth]{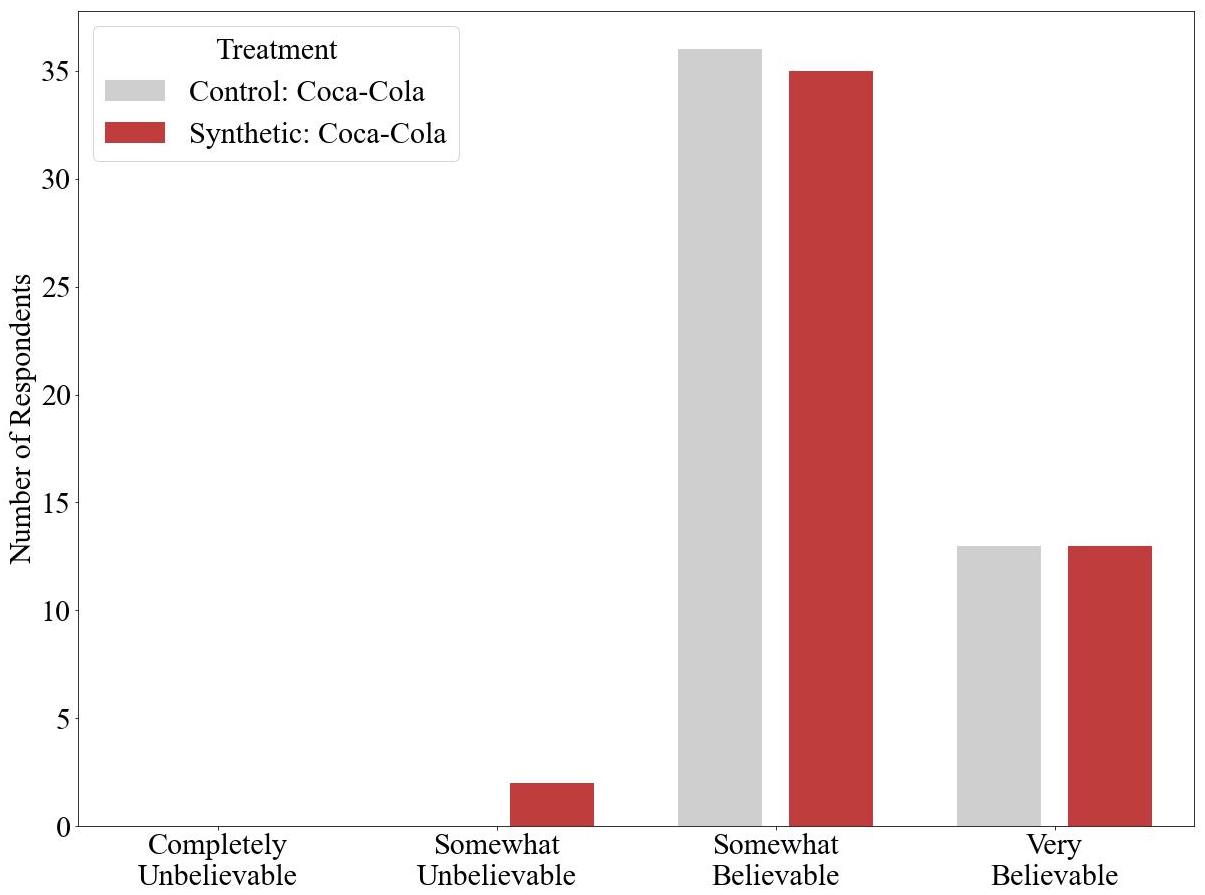} 
  \caption{Credibility assessment of treatment text (Q5).}
  \label{fig:credibility}
\end{figure}

\label{sec:headings}

\begin{table}
  \centering
  \begin{threeparttable}[c]
    \caption{Credibility Assessments Between Treatments (Raw Score)}
    \label{tab:three-part-table-cred}
    \begin{tabular}{ c c c c }
      \toprule
      Treatment & Mean & Diff. in Means & p-value\\
      \midrule
      Control & 3.265 & -- & -- \\
      \hline
      Synthetic & 3.220 & -0.045 & 0.638 \\
      \bottomrule
    \end{tabular}
    \begin{tablenotes}
      \item [a] {*}p < 0.05, {*}{*}p < 0.01, {*}{*}{*}p < 0.001.
    \end{tablenotes}
  \end{threeparttable}
\end{table}

The second method used the same binary aggregation used for Q3 (and originally proposed by Kreps et al.), bifurcating the Likert responses into two categories (``Unbelievable” and ``Believable”), and recording the average number of “Believable” responses. Once again, this numerical mean conveniently corresponded to the percentage of responses within each treatment group. The control group evaluated 100\% of treatment texts as credible under this binary threshold (mean: 1.000, SD: $\pm$ 0.000), while the synthetic treatment group evaluated 96\% of texts as credible (mean: 0.960, SD: $\pm$ 0.196). In other words, genuine and synthetic texts were virtually indistinguishable to survey participants through the lens of credibility, and rejection of the null hypothesis was once again statistically infeasible.

\begin{table}
  \centering
  \begin{threeparttable}[c]
    \caption{Credibility Assessments Between Treatments (Binary Score)}
    \label{tab:three-part-table-credbin}
    \begin{tabular}{ c c c c }
      \toprule
      Treatment & Mean & Diff. in Means & p-value\\
      \midrule
      Control & 1.000 & -- & -- \\
      \hline
      Synthetic & 0.960 & -0.040 & 0.159\\
      \bottomrule
    \end{tabular}
    \begin{tablenotes}
      \item [a] {*}p < 0.05, {*}{*}p < 0.01, {*}{*}{*}p < 0.001.
    \end{tablenotes}
  \end{threeparttable}
\end{table}

In order to begin exploring alternate hypothesis H2, the difference in means for Kreps et al.’s best-performing model (against a genuine world news article) was compared against the difference in means for this study’s synthetic experiment group (against a genuine Coca-Cola transcript). Conducting a t-test on this ``difference in difference-in-means" did not yield statistically significant results, though the markedly smaller deltas between this study (-0.040) and Kreps et al.’s study (0.130) suggests that the credibility gap between synthetic and genuine content may be narrowing. This gap reduction may be due to several factors, including the improvement of GPT-3.5 over GPT-2, the difference in text subject matter, or other miscellaneous reasons. However, conclusively isolating a cause lay beyond the scope of this experiment.

\begin{table}
  \centering
  \begin{threeparttable}[c]
    \caption{Credibility Assessments Between Text Type (Binary Score)}
    \label{tab:three-part-table-other}
    \begin{tabular}{ c c c c c c }
      \toprule
      & \multicolumn{1}{c}{\textit{Control}} & \multicolumn{1}{c}{\textit{Treatment}} & & & \\
      \cmidrule(l{20pt}r{20pt}){2-2} \cmidrule(l{20pt}r{20pt}){3-3}
      Study & \% ``Credible" & \% ``Credible" & Diff.\\
      \midrule
      Kreps & 0.71 & 0.84 & 0.13\\
      \hline
      Keeley & 1.00 & 0.96 & -0.04 \\
      \bottomrule
    \end{tabular}
    \begin{tablenotes}
      \item [a] {*}p < 0.05, {*}{*}p < 0.01, {*}{*}{*}p < 0.001.
    \end{tablenotes}
  \end{threeparttable}
\end{table}

\subsection{Limitations}

As in most studies, this experiment suffers from several limitations. First, critics might identify the difference in ``economic outlook" polarity between genuine and synthetic earnings call treatment texts as problematic. Because genuine earnings calls are usually framed in a ``positive light" (Estevez, 2022), and ChatGPT was instructed to generate calls with a “negative economic outlook,” the shift in investor sentiment witnessed in this study might – in the eyes of critics – lack profundity. 

In response, it first bears reiterating that these synthetic texts were generated autonomously and with minimal manual edits. In this context, the ability of a language model to produce convincing results (in less than 15 seconds) that effectively reverse the perceptions of experienced investors is impressive on its own. Nonetheless, two separate metrics were taken to more equitably assess influential power independent of positive or negative valence: the absolute value of short-term outlook score changes, and a binary outlook change metric. As seen in Section 4.1, these valence-agnostic metrics provided importance nuance to the raw scores of Q3.

Furthermore, a lexicon-based sentiment analysis was conducted on all three texts to assess how textual sentiment adjustment by GPT-3.5 may have influenced investor sentiment.\footnote{For clarification, two different kinds of “sentiments” are discussed here. Investor sentiment refers to belief about the future outlook of particular financial securities, which was the dependent variable this study sought to measure. Meanwhile, text sentiment refers to the calculation of a document’s tone by aggregating “positive” and “negative” words within the document. This analysis was carried out for the Coca-Cola control and treatment texts to more closely examine how ChatGPT manipulated text to influence investor belief.} The pre-trained “SnowyNLP" (Wang, 2013/2023) and “VADER” (Hutto \& Gilbert, 2014) sentiment models were used for the Chinese and English versions of each survey, respectively. Although the genuine Coca-Cola text did yield a slightly more positive compound sentiment score than the synthetic Coca-Cola text, the difference in sentiment scores was surprisingly small at 2.5\% for the Chinese version and 0.4\% for the English version (4.4\% and 14.8\% for HP). Thus, attributing change in investor sentiment based on text sentiment polarity alone seems far-fetched. Nonetheless, future studies seeking to provide a more balanced assessment of influential power would benefit by providing valence-equivalent control and treatment texts side-by-side for survey respondents. This would further isolate the influential effects of synthetic financial text against genuine financial text.

A second limitation of this study is its limited generalizability to other types of financial news, different investor audiences, alternative language models, and other publicly-listed companies. Earnings calls, for instance, might not be the main information source for some investors -- particularly outside of China. Global investors might also carry different biases and preexisting knowledge about equity markets across international boundaries. Because the ``average” investor comprises a majority of what we know as ``the market,” examining the impact of synthetic media on investors \textit{without} financial services backgrounds would be a valuable future research direction. Additionally, the use of GPT-3.5 was primarily driven by its rising popularity and public availability at the time of this writing. The comparison of influential power between different language models, particularly given the recent incorporation of GPT-4 into ChatGPT’s API (OpenAI, 2023), merits examination in future research, too. Lastly, this experiment was only able to control for and draw statistically conclusive results for financial text about The Coca-Cola Company. Although this study utilized a synthetic HP earnings call for exploratory comparison, any data involving HP -- other than perhaps comparisons between synthetic treatment results -- should be treated purely as exploratory. Future studies should absolutely analyze treatments from different industries to explore inter-industry differences more rigorously. 

Third, critics might point out that survey respondents don't have reason to doubt the credibility of a document provided in a survey. In other words, a certain amount of trust is embedded within the relationship between survey providers and survey respondents, helping explain the overwhelming confidence in text credibility across treatments. While this might be true, the decision to blind participants to the possibility of viewing misleading or synthetic text was deliberately taken for two important reasons. First, as seen via the Eli-Lilly Tweet episode described earlier, historical financial text manipulation (albeit non-synthetic) has often come under the guise of prominent enterprises. Therefore, replicating a ``real world" synthetic threat should not be any different. Second, previous studies (as mentioned in Chapter 2) suggest that human subjects perform no better than random guessing when they are instructed to classify synthetic versus genuine texts. In fact, one study even suggested that some survey participants were unresponsive to synthetic-text disclaimers (Kreps et al., 2022). Therefore, warning survey participants about potential exposure to synthetic or falsified texts might have led to skewed or diluted results.

Fourth, this experiment may have understated the ability of GPT-3.5 to persuade investors by virtue of its prompt. The prompt utilized in this experiment did not explicitly instruct ChatGPT to ``persuade.” Given preliminary research about the potency of ``prompt engineering” (Crothers et al., 2023), the absence of an instruction to persuade makes this study’s results all the more compelling. 

Finally, this study was carried out using simplified Chinese text, and GPT-3.5 is a language model primarily trained on English data. As seen in Guo et al.’s results, differences in human detection rates of synthetic media exist between Chinese and English text outputs (Guo et al., 2023). Given uniformly higher detection rates for synthetic financial texts produced in Chinese (Guo et al., 2023), this study expected to see declines in investor outlook changes and credibility assessments of synthetic treatment text in Chinese. Thus, replication of this study using English surveys and English-native participants would likely only strengthen the results found here.

\section{Conclusion}

\subsection{Overview}

To recap, this study sought to fill several gaps within the field of synthetic media literature. First, it highlighted the novelty of synthetic media threats versus traditional media threats in the face of growing public misunderstanding. 

Second, it showcased an asymmetrical focus within existing literature on audio-visual media, the impacts of synthetic media on political information environments, and Turing tests – particularly when juxtaposed against the popular emergence of another synthetic media modality: text. 

Third, it identified impact potential and resource limitations by organizing the synthetic media development process into a theoretical framework, pointing to financial information environments as prime candidates for future synthetic media disruption via text modalities. 

Finally, in the preceding chapter, it empirically tested the influential power of synthetic financial texts on investors compared to genuine financial texts. Rather than merely assessing the ability of survey participants to distinguish genuine from synthetic text, this study sought to go a step further; to measure the impact of synthetic media by observing its ability to manipulate \textit{belief}: the keystone linking information to action. Ultimately, these actions are responsible for the outcomes we fear as a society – whether they involve election tampering, or financial manipulation.

In this context, the results from this study’s experiment are concerning. As expected, the credibility ratings for genuine and synthetic texts were indistinguishably high – even for financial experts. Comparing these results against those in Kreps et al.’s study also suggests that the gap separating synthetic outputs and genuine media may be narrowing as generative models improve.  

More interestingly, however, falsified texts primed with a negative financial outlook significantly shifted investment behavior away from what it otherwise might have been under truthful information conditions. This behavior was evidenced across multiple variables related to stock market engagement, including share price predictions and willingness to invest. Alone, these drastic shifts in investor sentiment already paint a frightening picture for the financial world.  

However, examination of valence-agnostic metrics that ignored the positive and negative directionality of score changes revealed an even bleaker one. While comparing absolute value means between treatments lacked conclusive results, a statistically significant difference did exist for a binary “conversion” variable. One interpretation that reconciles these results is that genuine and synthetic texts tend to equivalently influence marginal improvements or degradations of an investor’s outlook on a company. For example, an investor that originally viewed Coca-Cola's short-term economic outlook as “Somewhat favorable” might change her outlook to “Very favorable” after reading a positively-primed, genuine earnings call transcript. However, synthetic earnings calls demonstrated a unique ability to influence \textit{extreme} changes in outlook about a company – converting investors from holding a positive outlook all the way across the spectrum to a negative outlook, and vice versa. In other words, synthetic financial media produced by ChatGPT may be better at changing minds than the truth is. This unique trend should inspire concern within the global financial community, particularly given the historical vulnerability of equity markets to investor sentiment shocks. 

Of course, the next steps forward involve understanding why this trend exists in order to determine what might be done about it. As mentioned in the “Limitations” section, sentiment analysis of the Chinese and English versions of each treatment text yielded disappointingly similar scores, discounting the probability that hyperbolic language within the synthetic treatment texts swayed survey respondents. Instead, reviewing survey participant write-in responses yielded more interesting insights into the ability of synthetic media to change investor sentiment.

\subsection{Potential Explanations}

\subsubsection{Write-In Response Analysis for Change in Economic Outlook (Q3)}

Post-treatment questions four and six asked survey participants to briefly explain their choices about adjusted short-term economic outlook (Q3) and credibility assessments (Q5), respectively. Responses to question four fell in several different categories; some participants latched on to key phrases, such as “worker strikes” \begin{CJK*}{UTF8}{gbsn}罢工\end{CJK*}, while others cited external economic factors and long-term company outlook. On the whole, however, investors overwhelmingly cited "metrics” as a key driver of adjusted short-term outlook -- across \textit{all} treatments. Specific metrics cited included “revenue” \begin{CJK*}{UTF8}{gbsn}收入\end{CJK*}, “operating profit” \begin{CJK*}{UTF8}{gbsn}营业利润\end{CJK*}, and “growth” \begin{CJK*}{UTF8}{gbsn}增长\end{CJK*}. For investors that changed their minds about Coca-Cola's short-term outlook, approximately 2 out of 3 cited data within the treatment text as a primary reason.\footnote{Entries that specifically mentioned “data,” “metrics,” financial figures (e.g. ``4\%"), or accounting terminology (e.g. “operating profit”) were tallied as ``metric" citations.} This estimate is also likely conservative, given that several investors within the “No change” category focused on profit declines in their written responses, but refused to update their pre-treatment score for their post-treatment response.

\subsubsection{Write-In Response Analysis for Credibility Assessments (Q5)}

Responses to question six provided even more context. One of the leading themes for high-credibility scores revolved around the use of ``data” \begin{CJK*}{UTF8}{gbsn}数据\end{CJK*} -- but particularly in an ``objective” \begin{CJK*}{UTF8}{gbsn}客观\end{CJK*} fashion. Although many participants showcased a belief that ``the data says everything” \begin{CJK*}{UTF8}{gbsn}数据说明一切\end{CJK*}, the balance of negative and positive self-critique by CEOs in the synthetic and genuine texts seemed to especially convince survey participants of the text’s authenticity. In the words of one participant, ``Who would splash dirty water on themselves?” \begin{CJK*}{UTF8}{gbsn}谁会往自己身上泼脏水\end{CJK*}? By contrast, the few texts that did receive low credibility scores were often cited as being ``exaggerated” \begin{CJK*}{UTF8}{gbsn}夸张\end{CJK*} or incorrect.

Evaluating the treatment text’s metric accuracy against Coca-Cola's actual earnings release in Q2 2018 validated the survey respondents’ focus on balanced metrics. As seen in Appendix C, ChatGPT’s output presented some figures accurately (e.g. ``net revenue of \$8.9 billion”), but still managed to spin some of them within a negative context (Q2 2018 Earnings Release, 2018). In the synthetic Coca-Cola transcript, many of the negative modifiers were also inaccurate. For instance, the transcript reported that ``operating income declined by 5\%” while the actual earnings release suggested an increase of 34\% (Q2 2018 Earnings Release, 2018). Interestingly, some negative modifiers were also incorrectly conservative. For instance, the ``2\% year-over-year" revenue decline reported by ChatGPT was actually a decline of 8\% (Q2 2018 Earnings Release, 2018). Whether the balance in these inconsistencies was random or deliberate is unclear and left to future research. 

A final, more concerning theme for credibility assessments centered on source trustworthiness. Dozens of respondents cited the CEO’s direct commentary, Coca-Cola's brand reputation, and/or the company’s public listing on equity markets as reasons for believing the content within the text. For example, one participant was quoted as saying, ``This was published by The Coca-Cola Company’s CEO, the company is prestigious, and it is a large brand, so [the text] is trustworthy” \begin{CJK*}{UTF8}{gbsn}因为这个是可口可乐公司CEO发布的公告,可口可乐在国际上还是比较享有信誉的,也是大品牌,所以信得过\end{CJK*}. Another respondent remarked, ``A company can’t really release fake news” \begin{CJK*}{UTF8}{gbsn}一个公司不太可能发布假消息\end{CJK*}. While these responses might partially reflect China's more controlled information environments, the responses also correspond with research that links perceptions of information authenticity with source trustworthiness (Hinkis \& Turner, 2022). 

\subsection{A Path Forward: Countermeasures}

Unfortunately, the identification of metrics as a key influential variable across \textit{all} treatments leaves the connection between synthetic financial text and outlook conversions a mystery. However, the responsiveness of survey participants to balanced metrics and source authenticity serves as an excellent starting point for future research – as well as for deliberating ways to mitigate the harmful effects of synthetic media. Currently, discussion of countermeasures within preexisting literature can be divided into several areas (Helmus, 2022; Hwang, 2020). Three overarching categories are considered in the following section: detection methods, media literacy campaigns, and policymaking.

\subsubsection{Detection}

Detection methods typically involve the use of machine learning or deep learning algorithms to identify synthetic material. From the perspective of the synthetic media development framework, these methods can be levied within the dissemination or consumption stages. Both private companies (e.g. Meta) and public organizations (e.g. The Defense Advanced Research Projects Agency) have made concerted efforts to develop these kinds of solutions (Helmus, 2022; Hwang, 2020) -- and across multiple modalities. For example, several common algorithms for identifying synthetic text include Grover, RoBERTa, and GLTR (Crothers et al., 2023; Pu et al., 2022; Guo et al., 2023). These models use frequency features, fluency features, and other types of variables for classifying genuine text against synthetic text (Crothers et al., 2023). Unfortunately, as discussed earlier, these efforts have tended to fall victim to the same generalization problem encountered during Meta’s DFDC (Helmus, 2022). Indeed, one recent study on text detection found that “open-domain detection schemes fail to generalize to \textit{in-the-wild} [sic] synthetic text, and that most defenses are not robust under adversarial settings” (Pu et al., 2022). 

Another approach that falls within the detection countermeasure category involves the manipulation of data to prevent its malign use. Several names exist for the outcomes of this process, to include “radioactive” (Helmus, 2022; Hwang, 2020) or “watermarked” data (A Watermark for LLMs - a Hugging Face Space by Tomg-Group-Umd, n.d.). Regardless of name, the premise remains the same: embed slight alterations (undetectable to the human eye) within public data likely to be used for training generative algorithms. This provides preemptive clues about malicious synthetic content generation (Helmus, 2022). Flagging synthetic content prior to consumption has also been empirically linked to limiting the spread of synthetic media (Helmus, 2022; Iacobucci et al., 2021). However, separate studies point to workarounds that algorithm developers can use to easily bypass these safeguards (Sadasivan et al., 2023). While privately safeguarding detection algorithms as national security products might benefit government stakeholders temporarily (Helmus, 2022), the general public would likely still suffer. 

In practice, it is possible that opensource detection methods will always run into the generalization problem. For this reason, private and institutional investors might consider disconnecting from high-volume, sentiment-sensitive information inputs like Twitter. Alternatively, if the financial services world insists on continuously developing detection algorithms, these products would benefit by identifying volatile changes in consumer sentiment, focusing on source authenticity, and flagging metric inaccuracies. One product that appears to be moving in the right direction is the DeepTrust algorithm, proposed by Pok Wah Chan in 2022. DeepTrust combines both machine learning \textit{and} natural language processing to classify anomalous market changes due to suspicious Tweets (Chan, 2022). 

\subsubsection{Media Literacy Campaigns}

Media literacy campaigns are efforts to educate people – in this case, private and institutional investors – about the threats that synthetic media pose. From the perspective of the synthetic media development framework, this method is primarily targeted at breaking the self-perpetuating, consumption-dissemination cycle. Several independent organizations currently undertake this initiative via educational materials, seminars, blogs, etc., including WITNESS (WITNESS Media Lab | Prepare, Don’t Panic: Synthetic Media and Deepfakes - WITNESS Media Lab, n.d.) and the Atlantic Council’s Digital Forensic Research Lab (“About the Atlantic Council,” n.d.). 

However, these organizations face a challenging paradox; although several studies prove that synthetic media awareness reduces both the likelihood of consumers to believe it (Shin \& Lee, 2022) and spread it (Iacobucci et al., 2021; Shin \& Lee, 2022), increased awareness also inflates peoples’ (mistaken) belief in their ability to detect it (Cochran \& Napshin, 2021). Implementing this balancing act in the global financial community might amount to cautioning private investors on social media investment forums, selectively publishing warnings through investment regulators such as the SEC, and adjusting new-hire training curricula to prevent the disruption of corporate investments. Currently, the SEC publishes a list of fraudulent companies on its website (SEC.Gov | Public Alert: Unregistered Soliciting Entities (PAUSE), n.d.), as well as investor training to avoid stock “pump and dump” schemes (Pump and Dump Schemes | Investor.Gov, n.d.). These resources do address investor vulnerability to source-trustworthiness biases, but might additionally benefit by including descriptions of synthetic threat technology, periodic updates on targeted industries, and warnings about threat actor approaches (e.g. metric balancing). 

\subsubsection{Policy}

Given the relatively nascent emergence of synthetic media, and the speed with which it is evolving, policymakers have only recently begun formulating solutions for mitigating its harmful effects. However, policymaking does arguably represent the most flexible category of countermeasure, capable of targeting any stage of the synthetic media development process.

For example, China has clearly focused on the production stage of synthetic media development, addressing the problem at its root. In November of 2022, the Cyberspace Administration of China (CAC) released its “Regulations on the management of the in-depth synthesis of Internet information services” (Cyberspace Administration of China \begin{CJK*}{UTF8}{gbsn}中华人民共和国国家互联网信息办公室\end{CJK*}, 2022). According to this policy, the burden of responsibility for dissemination and consumption of synthetic media material falls entirely on the shoulders of “synthetic content providers” \begin{CJK*}{UTF8}{gbsn}深度合成服务提供者\end{CJK*} -- to include managing data privacy, criminal reporting, safety audits, content management, and even rumor dispersion (Cyberspace Administration of China \begin{CJK*}{UTF8}{gbsn}中华人民共和国国家互联网信息办公室\end{CJK*}, 2022; China Law Translate, 2022). Following the production of several models mirroring ChatGPT in China, the CAC drafted a follow-on regulation on “generative artificial intelligence” that appears to impose more specific guidance, including preemptive government security reviews and data management guidelines (Cyberspace Administration of China\begin{CJK*}{UTF8}{gbsn}中华人民共和国国家互联网信息办公室\end{CJK*}, 2023; Coldewey, 2023)

By contrast, the European Union’s (EU) “Digital Services Act” has focused more on the dissemination stage of the synthetic media development framework. The Act, which was released in October of 2022, stipulates that social media platforms must empower users to file complaints, implement mechanisms for resolving grievances, and assign representative liaisons to EU member states (EUR-Lex - 32022R2065 - EN - EUR-Lex, 2022).  

Finally, the United States has predominantly targeted the consumption-dissemination cycle within the synthetic media development process -- primarily evidenced by its proposal of the “DEEP FAKES Accountability Act” in 2019 (116th Congress (2019-2020), 2019). This bill specifically criminalizes failure to adhere to “watermarking” standards for synthetically-generated content, assigning a maximum sentence of up to five years in prison (116th Congress (2019-2020), 2019). 

The decision regarding which stage to target will likely revolve around value prioritization and existing technology frameworks, which can differ greatly from country to country. For instance, China’s policy measure likely centers on synthetic content providers rather than individual criminals and foreign threats because it can afford to from a regulatory standpoint. Unlike many other countries, China’s "Great Firewall” is carefully monitored and insulated from foreign influence (Roberts, 2018). Meanwhile, the United States continuously struggles to balance the threat potential of synthetic disinformation against freedom of speech protections – even for “false speech” (Baek, 2020). Both nations likely face sacrificing innovative and benevolent synthetic media development in exchange for mitigating its negative effects. Ultimately, the varied approaches employed by policymakers around the world reaffirm the global reach of synthetic media influence, as well as the careful tradeoffs that must be made to control it. 

\subsection{Final Reflections}

Prior to November of 2022, the topic of synthetic media influence was largely buried within academic journals and constrained to isolated conversations about national security. The release of ChatGPT likely changed that forever. Since then, novel uses and adaptations of its underlying technology have been developed rapidly. For instance, within four months of ChatGPT’s inauguration (and following implementation of this experiment’s survey), OpenAI announced the release of GPT-4 (OpenAI, 2023). The announcement included claims of improved performance across multiple benchmarks, and has led to the adoption of GPT-4 by multiple organizations – including financial services giant Morgan Stanley (OpenAI, 2023). Additionally, global research teams of all sizes have begun scrambling to build language models that emulate ChatGPT’s success. For instance, a team from Fudan University in Shanghai announced the “MOSS” model in February of 2023, while Chinese technology giant Baidu released “Ernie Bot” in March (Yang, 2023). Finally, models that specialize in specific domains of day-to-day life have begun emerging as well. One such model, called “BloombergGPT,” was specifically designed to excel in financial matters, claiming improved performance across multiple tasks such as named entity recognition and sentiment analysis (Wu et al., 2023). Although studies suggest that increasingly large language models will experience diminishing marginal returns at influencing variables like ``credibility” (Kreps et al., 2022), it will be interesting to see whether such models deter or amplify market vulnerability alongside a synthetic media influx.  

On a final note, this study’s use of Chinese survey participants to evaluate the deployment of an American language model was admittedly a product of happenstance. However, it also serves as an apt reminder (much like the Great Recession of 2008) that the financial community is a global one; the fates of its stakeholders – especially China and the United States – are closely interwoven. Safeguarding the security of financial information environments cannot be the sole burden of any one entity, and may very well serve as a proving ground for international cooperation in the future. 

\nocite{*} 

\bibliographystyle{apalike}  
\bibliography{main}  

\begin{thebibliography}{}

\bibitem[noa, a]{noauthor_about_nodate}
About the {Atlantic} {Council}.

\bibitem[noa, b]{noauthor_data_nodate}
Data {Science} {Pipeline}.

\bibitem[noa, c]{noauthor_new_nodate}
New {GPT}-3 capabilities: {Edit} \& insert.

\bibitem[noa, d]{noauthor_pump_nodate}
Pump and {Dump} {Schemes} {\textbar} {Investor}.gov.

\bibitem[noa, e]{noauthor_secgov_nodate}
{SEC}.gov {\textbar} {Financial} {Reporting} {Manual}.

\bibitem[noa, f]{noauthor_secgov_nodate-1}
{SEC}.gov {\textbar} {Public} {Alert}: {Unregistered} {Soliciting} {Entities}
  ({PAUSE}).

\bibitem[noa, g]{noauthor_sp_nodate}
S\&{P} {Global} 100.

\bibitem[noa, h]{noauthor_sp_nodate-1}
S\&{P} {Global} 100 {Constituents}.

\bibitem[noa, i]{noauthor_watermark_nodate}
A {Watermark} for {LLMs} - a {Hugging} {Face} {Space} by tomg-group-umd.

\bibitem[noa, j]{noauthor_why_nodate}
Why doesn't {ChatGPT} know about {X}? {\textbar} {OpenAI} {Help} {Center}.

\bibitem[noa, k]{noauthor_witness_nodate}
{WITNESS} {Media} {Lab} {\textbar} {Prepare}, {Don}'t {Panic}: {Synthetic}
  {Media} and {Deepfakes} - {WITNESS} {Media} {Lab}.

\bibitem[noa, 2008]{noauthor_digital_2008}
 (2008).
\newblock Digital {Forensics}: {Photo} {Tampering} {Throughout} {History}
  [{Slide} {Show}].

\bibitem[noa, 2013]{noauthor_coca-cola_2013}
 (2013).
\newblock Coca-{Cola} says to invest over \$4 billion in {China} in 2015-2017
  {\textbar} {Reuters}.

\bibitem[noa, 2018a]{noauthor_bloomberg_2018}
 (2018a).
\newblock Bloomberg {Launches} a {Twitter} {Feed} {Optimized} for {Trading}
  {\textbar} {Press} {\textbar} {Bloomberg} {LP}.
\newblock {\em Bloomberg L.P.}

\bibitem[noa, 2018b]{noauthor_coca-cola_2018}
 (2018b).
\newblock Coca-{Cola} {Company} ({The}) ({KO}) {Q2} 2018 {Earnings}
  {Conference} {Call} {Transcript}.
\newblock Section: earnings.

\bibitem[noa, 2018c]{noauthor_q2_2018}
 (2018c).
\newblock Q2 2018 {Earnings} {Release}.

\bibitem[noa, 2019]{noauthor_aiaaic_2019}
 (2019).
\newblock {AIAAIC} - {AI} and algorithmic incidents and controversies.

\bibitem[noa, 2020a]{noauthor_deepfake_2020}
 (2020a).
\newblock Deepfake {Detection} {Challenge} {Results}: {An} open initiative to
  advance {AI}.

\bibitem[noa, 2020b]{noauthor_robot_2020}
 (2020b).
\newblock A robot wrote this entire article. {Are} you scared yet, human?
\newblock {\em The Guardian}.

\bibitem[noa, 2021]{noauthor_ethiopias_2021}
 (2021).
\newblock Ethiopia's warring sides locked in disinformation battle.

\bibitem[noa, 2022a]{noauthor_academic_2022}
 (2022a).
\newblock Academic {Webinar}: {AI} {Military} {Innovation} and {U}.{S}.
  {Defense} {Strategy}.

\bibitem[noa, 2022b]{noauthor_eur-lex_2022}
 (2022b).
\newblock {EUR}-{Lex} - {32022R2065} - {EN} - {EUR}-{Lex}.
\newblock Doc ID: 32022R2065 Doc Sector: 3 Doc Title: Regulation (EU) 2022/2065
  of the European Parliament and of the Council of 19 October 2022 on a Single
  Market For Digital Services and amending Directive 2000/31/EC (Digital
  Services Act) (Text with EEA relevance) Doc Type: R Usr\_lan: en.

\bibitem[noa, 2022c]{noauthor_how_2022}
 (2022c).
\newblock How a {Mysterious} {China} {Screenshot} {Spurred} \$450 {Billion}
  {Rally}.
\newblock {\em Bloomberg.com}.

\bibitem[noa, 2022d]{noauthor_introducing_2022}
 (2022d).
\newblock Introducing {ChatGPT}.

\bibitem[noa, 2022e]{noauthor_regulating_2022}
 (2022e).
\newblock Regulating tech use to curb 'deepfake' risks.

\bibitem[noa, 2022f]{noauthor_transcript_2022}
 (2022f).
\newblock Transcript - {Why} the {Zelensky} {Deepfake} {Failed}.

\bibitem[noa, 2023]{noauthor_ai_2023}
 (2023).
\newblock {AI} {Index} {Report} 2023 – {Artificial} {Intelligence} {Index}.

\bibitem[116th Congress (2019-2020),
  2019a]{116th_congress_2019-2020_hr3230_2019}
116th Congress (2019-2020) (2019a).
\newblock H.{R}.3230 - 116th {Congress} (2019-2020): {DEEP} {FAKES}
  {Accountability} {Act} {\textbar} {Congress}.gov {\textbar} {Library} of
  {Congress}.

\bibitem[116th Congress (2019-2020), 2019b]{116th_congress_2019-2020_s_2019}
116th Congress (2019-2020) (2019b).
\newblock S. {Rept}. 116-93 - {DEEPFAKE} {REPORT} {ACT} {OF} 2019 {\textbar}
  {Congress}.gov {\textbar} {Library} of {Congress}.

\bibitem[Alba, 2022]{alba_twitter_2022}
Alba, D. (2022).
\newblock Twitter {Pauses} \$8 {Subscription} {Program} {After} {Fake}
  {Accounts} {Proliferate}.
\newblock {\em Bloomberg.com}.

\bibitem[Allyn, 2022]{allyn_deepfake_2022}
Allyn, B. (2022).
\newblock Deepfake video of {Zelenskyy} could be 'tip of the iceberg' in info
  war, experts warn.
\newblock {\em NPR}.

\bibitem[Baek, 2020]{baek_free_2020}
Baek, S. (2020).
\newblock Free {Speech} in the {Digital} {Age}: {Deepfakes} and the
  {Marketplace} of {Ideas}.

\bibitem[Baker and Wurgler, 2007]{baker_investor_2007}
Baker, M. and Wurgler, J. (2007).
\newblock Investor {Sentiment} in the {Stock} {Market}.
\newblock {\em The Journal of Economic Perspectives}, 21(2):129--151.
\newblock Publisher: American Economic Association.

\bibitem[Bateman, 2020]{bateman_deepfakes_2020}
Bateman, J. (2020).
\newblock Deepfakes and {Synthetic} {Media} in the {Financial} {System}:
  {Assessing} {Threat} {Scenarios}.
\newblock {\em Carnegie Endowment for International Peace}.

\bibitem[Beresneva, 2016]{beresneva_computer-generated_2016}
Beresneva, D. (2016).
\newblock Computer-{Generated} {Text} {Detection} {Using} {Machine} {Learning}:
  {A} {Systematic} {Review}.
\newblock In Métais, E., Meziane, F., Saraee, M., Sugumaran, V., and Vadera,
  S., editors, {\em Natural {Language} {Processing} and {Information}
  {Systems}}, Lecture {Notes} in {Computer} {Science}, pages 421--426, Cham.
  Springer International Publishing.

\bibitem[BM, 2022]{bm_what_2022}
BM, N. (2022).
\newblock What is an encoder decoder model?

\bibitem[Bray et~al., 2022]{bray_testing_2022}
Bray, S.~D., Johnson, S.~D., and Kleinberg, B. (2022).
\newblock Testing {Human} {Ability} {To} {Detect} {Deepfake} {Images} of
  {Human} {Faces}.
\newblock arXiv:2212.05056 [cs].

\bibitem[Brewster, 2021]{brewster_fraudsters_2021}
Brewster, T. (2021).
\newblock Fraudsters {Cloned} {Company} {Director}’s {Voice} {In} \$35
  {Million} {Bank} {Heist}, {Police} {Find}.
\newblock Section: Cybersecurity.

\bibitem[Brown et~al., 2020]{brown_language_2020}
Brown, T.~B., Mann, B., Ryder, N., Subbiah, M., Kaplan, J., Dhariwal, P.,
  Neelakantan, A., Shyam, P., Sastry, G., Askell, A., Agarwal, S.,
  Herbert-Voss, A., Krueger, G., Henighan, T., Child, R., Ramesh, A., Ziegler,
  D.~M., Wu, J., Winter, C., Hesse, C., Chen, M., Sigler, E., Litwin, M., Gray,
  S., Chess, B., Clark, J., Berner, C., McCandlish, S., Radford, A., Sutskever,
  I., and Amodei, D. (2020).
\newblock Language {Models} are {Few}-{Shot} {Learners}.
\newblock arXiv:2005.14165 [cs].

\bibitem[{BuzzFeedVideo}, 2018]{buzzfeedvideo_you_2018}
{BuzzFeedVideo} (2018).
\newblock You {Won}’t {Believe} {What} {Obama} {Says} {In} {This} {Video}!
  ;).

\bibitem[Caldarelli et~al., 2020]{caldarelli_role_2020}
Caldarelli, G., De~Nicola, R., Del~Vigna, F., Petrocchi, M., and Saracco, F.
  (2020).
\newblock The role of bot squads in the political propaganda on {Twitter}.
\newblock {\em Communications Physics}, 3(1):1--15.
\newblock Number: 1 Publisher: Nature Publishing Group.

\bibitem[Cao et~al., 2023]{cao_comprehensive_2023}
Cao, Y., Li, S., Liu, Y., Yan, Z., Dai, Y., Yu, P.~S., and Sun, L. (2023).
\newblock A {Comprehensive} {Survey} of {AI}-{Generated} {Content} ({AIGC}):
  {A} {History} of {Generative} {AI} from {GAN} to {ChatGPT}.
\newblock arXiv:2303.04226 [cs].

\bibitem[Chan, 2022]{chan_deeptrust_2022}
Chan, P.~W. (2022).
\newblock {DeepTrust}: {A} {Reliable} {Financial} {Knowledge} {Retrieval}
  {Framework} {For} {Explaining} {Extreme} {Pricing} {Anomalies}.
\newblock arXiv:2203.08144 [cs, q-fin].

\bibitem[Chesney and Citron, 2019]{chesney_deep_2019}
Chesney, B. and Citron, D. (2019).
\newblock Deep {Fakes}: {A} {Looming} {Challenge} for {Privacy}, {Democracy},
  and {National} {Security}.
\newblock {\em CALIFORNIA LAW REVIEW}, 107.

\bibitem[{China Law Translate}, 2022]{china_law_translate__2022}
{China Law Translate} (2022).
\newblock Provisions on the administration of deep synthesis internet
  information services (draft for solicitation of comments).

\bibitem[Cochran and Napshin, 2021]{cochran_deepfakes_2021}
Cochran, J.~D. and Napshin, S.~A. (2021).
\newblock Deepfakes: {Awareness}, {Concerns}, and {Platform} {Accountability}.
\newblock {\em Cyberpsychology, Behavior, and Social Networking},
  24(3):164--172.
\newblock Publisher: Mary Ann Liebert, Inc., publishers.

\bibitem[Coldewey, 2023]{coldewey_prohibition_2023}
Coldewey, D. (2023).
\newblock Prohibition of {AI} that 'subverts state power' in {China} may chill
  its nascent industry.

\bibitem[Crothers et~al., 2023]{crothers_machine_2023}
Crothers, E., Japkowicz, N., and Viktor, H. (2023).
\newblock Machine {Generated} {Text}: {A} {Comprehensive} {Survey} of {Threat}
  {Models} and {Detection} {Methods}.
\newblock arXiv:2210.07321 [cs].

\bibitem[Damiani, 2019]{damiani_voice_2019}
Damiani, J. (2019).
\newblock A {Voice} {Deepfake} {Was} {Used} {To} {Scam} {A} {CEO} {Out} {Of}
  \$243,000.

\bibitem[DiResta, 2020]{diresta_supply_2020}
DiResta, R. (2020).
\newblock The {Supply} of {Disinformation} {Will} {Soon} {Be} {Infinite}.
\newblock Section: Ideas.

\bibitem[Du et~al., 2020]{du_towards_2020}
Du, M., Pentyala, S., Li, Y., and Hu, X. (2020).
\newblock Towards {Generalizable} {Deepfake} {Detection} with {Locality}-aware
  {AutoEncoder}.
\newblock arXiv:1909.05999 [cs].

\bibitem[Estevez, 2022]{estevez_when_2022}
Estevez, E. (2022).
\newblock When {Must} a {Company} {Announce} {Earnings}?

\bibitem[Gregory, 2021]{gregory_world_2021}
Gregory, S. (2021).
\newblock The {World} {Needs} {Deepfake} {Experts} to {Stem} {This} {Chaos}
  {\textbar} {WIRED}.

\bibitem[Guo, 2023]{guo_guo_2023}
Guo, B. (2023).
\newblock "{Guo} {Biyang} {Interview} - 27 {February} 2023".

\bibitem[Guo et~al., 2023]{guo_how_2023}
Guo, B., Zhang, X., Wang, Z., Jiang, M., Nie, J., Ding, Y., Yue, J., and Wu, Y.
  (2023).
\newblock How {Close} is {ChatGPT} to {Human} {Experts}? {Comparison} {Corpus},
  {Evaluation}, and {Detection}.
\newblock arXiv:2301.07597 [cs].

\bibitem[Hameleers et~al., 2022]{hameleers_you_2022}
Hameleers, M., van~der Meer, T. G. L.~A., and Dobber, T. (2022).
\newblock You {Won}’t {Believe} {What} {They} {Just} {Said}! {The} {Effects}
  of {Political} {Deepfakes} {Embedded} as {Vox} {Populi} on {Social} {Media}.
\newblock {\em Social Media + Society}, 8(3):20563051221116346.
\newblock Publisher: SAGE Publications Ltd.

\bibitem[Harwell, 2022]{harwell_fake_2022}
Harwell, D. (2022).
\newblock A fake tweet sparked panic at {Eli} {Lilly} and may have cost
  {Twitter} millions.
\newblock {\em Washington Post}.

\bibitem[Helmus, 2022]{helmus_artificial_2022}
Helmus, T.~C. (2022).
\newblock Artificial {Intelligence}, {Deepfakes}, and {Disinformation}: {A}
  {Primer}.
\newblock Technical report, RAND Corporation.

\bibitem[Hiebert, 2022]{hiebert_democracies_2022}
Hiebert, K. (2022).
\newblock Democracies {Are} {Dangerously} {Unprepared} for {Deepfakes}.

\bibitem[Hinkis and Turner, 2022]{hinkis_chinese_2022}
Hinkis, N. and Turner, L. (2022).
\newblock Chinese {State} {Media}’s {Global} {Influencer} {Operation}: {Why}
  {It} {Matters}.

\bibitem[Horwitz, 2019]{horwitz_chinese_2019}
Horwitz, J. (2019).
\newblock Chinese face-swapping app goes viral, sparks privacy concerns.
\newblock {\em Reuters}.

\bibitem[Hsu and Thompson, 2023]{hsu_disinformation_2023}
Hsu, T. and Thompson, S.~A. (2023).
\newblock Disinformation {Researchers} {Raise} {Alarms} {About} {A}.{I}.
  {Chatbots}.
\newblock {\em The New York Times}.

\bibitem[Hu, 2023]{hu_chatgpt_2023}
Hu, K. (2023).
\newblock {ChatGPT} sets record for fastest-growing user base - analyst note.
\newblock {\em Reuters}.

\bibitem[Huang, 2023]{huang_alarmed_2023}
Huang, K. (2023).
\newblock Alarmed by {A}.{I}. {Chatbots}, {Universities} {Start} {Revamping}
  {How} {They} {Teach} - {The} {New} {York} {Times}.

\bibitem[Hutto and Gilbert, 2014]{hutto_vader_2014}
Hutto, C. and Gilbert, E. (2014).
\newblock {VADER}: {A} {Parsimonious} {Rule}-{Based} {Model} for {Sentiment}
  {Analysis} of {Social} {Media} {Text}.
\newblock {\em Proceedings of the International AAAI Conference on Web and
  Social Media}, 8(1):216--225.
\newblock Number: 1.

\bibitem[Hwang, 2020]{hwang_deepfakes_2020}
Hwang, T. (2020).
\newblock Deepfakes: {A} {Grounded} {Threat} {Assessment}.

\bibitem[Iacobucci et~al., 2021]{iacobucci_deepfakes_2021}
Iacobucci, S., De~Cicco, R., Michetti, F., Palumbo, R., and Pagliaro, S.
  (2021).
\newblock Deepfakes {Unmasked}: {The} {Effects} of {Information} {Priming} and
  {Bullshit} {Receptivity} on {Deepfake} {Recognition} and {Sharing}
  {Intention}.
\newblock {\em Cyberpsychology, Behavior and Social Networking},
  24(3):194--202.

\bibitem[Johnston, 2022]{johnston_chinas_2022}
Johnston (2022).
\newblock China’s {Stock} {Markets} vs. {U}.{S}. {Stock} {Markets}.

\bibitem[Jones, 2001]{jones_natural_2001}
Jones, K.~S. (2001).
\newblock Natural language processing: a historical review.
\newblock {\em University of Cambridge}.

\bibitem[Kenton, 2022]{kenton_what_2022}
Kenton, W. (2022).
\newblock What {Is} a {Bloomberg} {Terminal}? {Functions}, {Costs}, and
  {Alternatives}.

\bibitem[Kirchner et~al., 2023]{kirchner_new_2023}
Kirchner, J.~H., Ahmad, L., Aaronson, S., and Leike, J. (2023).
\newblock New {AI} classifier for indicating {AI}-written text.

\bibitem[Korshunov and Marcel, 2020]{korshunov_deepfake_2020}
Korshunov, P. and Marcel, S. (2020).
\newblock Deepfake detection: humans vs. machines.
\newblock arXiv:2009.03155 [cs, eess].

\bibitem[Kreps et~al., 2022]{kreps_all_2022}
Kreps, S., McCain, R.~M., and Brundage, M. (2022).
\newblock All the {News} {That}’s {Fit} to {Fabricate}: {AI}-{Generated}
  {Text} as a {Tool} of {Media} {Misinformation}.
\newblock {\em Journal of Experimental Political Science}, 9(1):104--117.
\newblock Publisher: Cambridge University Press.

\bibitem[Kurenkov, 2022]{kurenkov_lessons_2022}
Kurenkov, A. (2022).
\newblock Lessons from the {GPT}-4chan {Controversy}.

\bibitem[Lee and Shin, 2022]{lee_something_2022}
Lee, J. and Shin, S.~Y. (2022).
\newblock Something that {They} {Never} {Said}: {Multimodal} {Disinformation}
  and {Source} {Vividness} in {Understanding} the {Power} of {AI}-{Enabled}
  {Deepfake} {News}.
\newblock {\em Media Psychology}, 25(4):531--546.
\newblock Publisher: Routledge \_eprint:
  https://doi.org/10.1080/15213269.2021.2007489.

\bibitem[Leswing, 2023]{leswing_microsofts_2023}
Leswing, K. (2023).
\newblock Microsoft's {Bing} {A}.{I}. made several factual errors in last
  week's launch demo.

\bibitem[Locke, 1982]{locke_beliefs_1982}
Locke, D. (1982).
\newblock Beliefs, {Desires} and {Reasons} for {Action}.
\newblock {\em American Philosophical Quarterly}, 19(3):241--249.
\newblock Publisher: [North American Philosophical Publications, University of
  Illinois Press].

\bibitem[Lohn and Musser, 2022]{lohn_ai_2022}
Lohn, A. and Musser, M. (2022).
\newblock {AI} and {Compute}.

\bibitem[Lyons, 2020]{lyons_indian_2020}
Lyons, K. (2020).
\newblock An {Indian} politician used {AI} to translate his speech into other
  languages to reach more voters.

\bibitem[Ma, 2023]{ma_hp_2023}
Ma, S. (2023).
\newblock {HP} sees long-term benefits in {Chinese} businesses.

\bibitem[Mac et~al., 2022]{mac_verifiable_2022}
Mac, R., Mullin, B., Conger, K., and Isaac, M. (2022).
\newblock A {Verifiable} {Mess}: {Twitter} {Users} {Create} {Havoc} by
  {Impersonating} {Brands}.
\newblock {\em The New York Times}.

\bibitem[Mai et~al., 2023]{mai_warning_2023}
Mai, K.~T., Bray, S.~D., Davies, T., and Griffin, L.~D. (2023).
\newblock Warning: {Humans} {Cannot} {Reliably} {Detect} {Speech} {Deepfakes}.
\newblock arXiv:2301.07829 [cs, eess].

\bibitem[Mehdi, 2023]{mehdi_reinventing_2023}
Mehdi, Y. (2023).
\newblock Reinventing search with a new {AI}-powered {Microsoft} {Bing} and
  {Edge}, your copilot for the web.

\bibitem[{Mikael Thalen [@MikaelThalen]},
  2022]{mikael_thalen_mikaelthalen_deepfake_2022}
{Mikael Thalen [@MikaelThalen]} (2022).
\newblock A deepfake of {Ukrainian} {President} {Volodymyr} {Zelensky} calling
  on his soldiers to lay down their weapons was reportedly uploaded to a hacked
  {Ukrainian} news website today, per @{Shayan86} https://t.co/{tXLrYECGY4}.

\bibitem[Morrish, 2023]{morrish_fact-checkers_2023}
Morrish, L. (2023).
\newblock Fact-{Checkers} {Are} {Scrambling} to {Fight} {Disinformation} {With}
  {AI}.
\newblock {\em Wired}.
\newblock Section: tags.

\bibitem[Müller et~al., 2022]{muller_does_2022}
Müller, N.~M., Czempin, P., Dieckmann, F., Froghyar, A., and Böttinger, K.
  (2022).
\newblock Does {Audio} {Deepfake} {Detection} {Generalize}?
\newblock arXiv:2203.16263 [cs, eess].

\bibitem[Nguyen et~al., 2022]{nguyen_deep_2022}
Nguyen, T.~T., Nguyen, Q. V.~H., Nguyen, D.~T., Nguyen, D.~T., Huynh-The, T.,
  Nahavandi, S., Nguyen, T.~T., Pham, Q.-V., and Nguyen, C.~M. (2022).
\newblock Deep {Learning} for {Deepfakes} {Creation} and {Detection}: {A}
  {Survey}.
\newblock {\em Computer Vision and Image Understanding}, 223:103525.
\newblock arXiv:1909.11573 [cs, eess].

\bibitem[Nuñez, 2020]{nunez_snapchat_2020}
Nuñez, M. (2020).
\newblock Snapchat and {TikTok} {Embrace} ‘{Deepfake}’ {Video} {Technology}
  {Even} {As} {Facebook} {Shuns} {It}.
\newblock Section: Social Media.

\bibitem[of~China, 2022]{_-_2022}
of~China, C.~A. (2022).
\newblock "regulations on the administration of deep synthesis of internet
  information services" answers reporters' questions.

\bibitem[of~China, 2023]{_-_2023}
of~China, C.~A. (2023).
\newblock Notice of the national internet information office on the measures
  for the management of generating artificial intelligence services (draft for
  comments).

\bibitem[OpenAI, ]{openai_openai_nodate}
OpenAI.
\newblock {OpenAI} {API}.

\bibitem[OpenAI, 2022]{openai_chatgpt_2022}
OpenAI (2022).
\newblock {ChatGPT}.

\bibitem[OpenAI, 2023]{openai_gpt-4_2023}
OpenAI (2023).
\newblock {GPT}-4.

\bibitem[Phillips and Lorenz, 2021]{phillips_dumb_2021}
Phillips, M. and Lorenz, T. (2021).
\newblock ‘{Dumb} {Money}’ {Is} on {GameStop}, and {It}’s {Beating}
  {Wall} {Street} at {Its} {Own} {Game}.
\newblock {\em The New York Times}.

\bibitem[Pu et~al., 2022]{pu_deepfake_2022}
Pu, J., Sarwar, Z., Abdullah, S.~M., Rehman, A., Kim, Y., Bhattacharya, P.,
  Javed, M., and Viswanath, B. (2022).
\newblock Deepfake {Text} {Detection}: {Limitations} and {Opportunities}.
\newblock arXiv:2210.09421 [cs].

\bibitem[Roberts, 2018]{roberts_censored_2018}
Roberts, M. (2018).
\newblock {\em Censored: {A} {Distraction} and {Diversion} {Inside} {China}'s
  {Great} {Firewall}}.
\newblock Princeton University Press.

\bibitem[Roose, 2022]{roose_brilliance_2022}
Roose, K. (2022).
\newblock The {Brilliance} and {Weirdness} of {ChatGPT}.
\newblock {\em The New York Times}.

\bibitem[Roose, 2023]{roose_why_2023}
Roose, K. (2023).
\newblock Why a {Conversation} {With} {Bing}’s {Chatbot} {Left} {Me} {Deeply}
  {Unsettled} - {The} {New} {York} {Times}.

\bibitem[Ruby, 2023]{ruby_how_2023}
Ruby, M. (2023).
\newblock How {ChatGPT} {Works}: {The} {Models} {Behind} {The} {Bot}.

\bibitem[Sadasivan et~al., 2023]{sadasivan_can_2023}
Sadasivan, V.~S., Kumar, A., Balasubramanian, S., Wang, W., and Feizi, S.
  (2023).
\newblock Can {AI}-{Generated} {Text} be {Reliably} {Detected}?
\newblock arXiv:2303.11156 [cs].

\bibitem[{Sensity}, 2019]{sensity_reports_2019}
{Sensity} (2019).
\newblock Reports.

\bibitem[Shao et~al., 2018]{shao_spread_2018}
Shao, C., Ciampaglia, G.~L., Varol, O., Yang, K.-C., Flammini, A., and Menczer,
  F. (2018).
\newblock The spread of low-credibility content by social bots.
\newblock {\em Nature Communications}, 9(1):4787.
\newblock Number: 1 Publisher: Nature Publishing Group.

\bibitem[Shin and Yi, 2022]{shin_south_2022}
Shin, H. and Yi, H.~Y. (2022).
\newblock South {Korea} candidates woo young voters with 'deepfakes,' hair
  insurance.
\newblock {\em Reuters}.

\bibitem[Shin and Lee, 2022]{shin_effect_2022}
Shin, S.~Y. and Lee, J. (2022).
\newblock The {Effect} of {Deepfake} {Video} on {News} {Credibility} and
  {Corrective} {Influence} of {Cost}-{Based} {Knowledge} about {Deepfakes}.
\newblock {\em Digital Journalism}, 10(3):412--432.
\newblock Publisher: Routledge \_eprint:
  https://doi.org/10.1080/21670811.2022.2026797.

\bibitem[Simonite, 2019]{simonite_see_2019}
Simonite, T. (2019).
\newblock To {See} the {Future} of {Disinformation}, {You} {Build}
  {Robo}-{Trolls} {\textbar} {WIRED}.

\bibitem[Simonite, 2022]{simonite_zelensky_2022}
Simonite, T. (2022).
\newblock A {Zelensky} {Deepfake} {Was} {Quickly} {Defeated}. {The} {Next}
  {One} {Might} {Not} {Be} {\textbar} {WIRED}.

\bibitem[Somers, 2020]{somers_deepfakes_2020}
Somers, M. (2020).
\newblock Deepfakes, explained.

\bibitem[Stokel-Walker, 2019]{stokel-walker_its_2019}
Stokel-Walker, C. (2019).
\newblock It's not {Zao} app's privacy issues that you should be worried about.
\newblock {\em Wired UK}.
\newblock Section: tags.

\bibitem[van Minnen et~al., 2022]{van_minnen_initial_2022}
van Minnen, A., Ter~Heide, F. J.~J., Koolstra, T., de~Jongh, A., Karaoglu, S.,
  and Gevers, T. (2022).
\newblock Initial development of perpetrator confrontation using deepfake
  technology in victims with sexual violence-related {PTSD} and moral injury.
\newblock {\em Frontiers in Psychiatry}, 13:882957.

\bibitem[Vaswani et~al., 2017]{vaswani_attention_2017}
Vaswani, A., Shazeer, N., Parmar, N., Uszkoreit, J., Jones, L., Gomez, A.~N.,
  Kaiser, L., and Polosukhin, I. (2017).
\newblock Attention {Is} {All} {You} {Need}.
\newblock arXiv:1706.03762 [cs].

\bibitem[{Vecanoi}, 2021]{vecanoi_very_2021}
{Vecanoi} (2021).
\newblock Very realistic {Tom} {Cruise} {Deepfake} {\textbar} {AI} {Tom}
  {Cruise}.

\bibitem[Verdoliva, 2020]{verdoliva_media_2020}
Verdoliva, L. (2020).
\newblock Media {Forensics} and {DeepFakes}: an overview.
\newblock arXiv:2001.06564 [cs].

\bibitem[Vincent, 2018]{vincent_watch_2018}
Vincent, J. (2018).
\newblock Watch {Jordan} {Peele} use {AI} to make {Barack} {Obama} deliver a
  {PSA} about fake news - {The} {Verge}.

\bibitem[Vosoughi et~al., 2018]{vosoughi_spread_2018}
Vosoughi, S., Roy, D., and Aral, S. (2018).
\newblock The spread of true and false news online.
\newblock {\em Science}, 359(6380):1146--1151.
\newblock Publisher: American Association for the Advancement of Science.

\bibitem[Wakefield, 2022]{wakefield_deepfake_2022}
Wakefield, J. (2022).
\newblock Deepfake presidents used in {Russia}-{Ukraine} war.
\newblock {\em BBC News}.

\bibitem[Wang, 2023]{wang_snownlp_2023}
Wang, R. (2023).
\newblock {SnowNLP}: {Simplified} {Chinese} {Text} {Processing}.
\newblock original-date: 2013-11-26T11:46:56Z.

\bibitem[Warren, 2023]{warren_microsofts_2023}
Warren, T. (2023).
\newblock Microsoft’s {Bing} {AI}, like {Google}’s, also made dumb mistakes
  during first demo - {The} {Verge}.

\bibitem[Weiss, 2019]{weiss_deepfake_2019}
Weiss, M. (2019).
\newblock Deepfake {Bot} {Submissions} to {Federal} {Public} {Comment}
  {Websites} {Cannot} {Be} {Distinguished} from {Human} {Submissions}.
\newblock {\em Technology Science}.

\bibitem[Wiederhold, 2021]{wiederhold_can_2021}
Wiederhold, B.~K. (2021).
\newblock Can {Deepfakes} {Improve} {Therapy}?
\newblock {\em Cyberpsychology, Behavior, and Social Networking},
  24(3):147--148.
\newblock Publisher: Mary Ann Liebert, Inc., publishers.

\bibitem[Wu et~al., 2023]{wu_bloomberggpt_2023}
Wu, S., Irsoy, O., Lu, S., Dabravolski, V., Dredze, M., Gehrmann, S., Kambadur,
  P., Rosenberg, D., and Mann, G. (2023).
\newblock {BloombergGPT}: {A} {Large} {Language} {Model} for {Finance}.
\newblock arXiv:2303.17564 [cs, q-fin].

\bibitem[Yang, 2023]{yang_chinese_2023}
Yang, Z. (2023).
\newblock Chinese tech giant {Baidu} just released its answer to {ChatGPT}.

\bibitem[Zellers, 2019]{zellers_counteracting_2019}
Zellers, R. (2019).
\newblock Counteracting neural disinformation with {Grover}.

\bibitem[Zellers et~al., 2020]{zellers_defending_2020}
Zellers, R., Holtzman, A., Rashkin, H., Bisk, Y., Farhadi, A., Roesner, F., and
  Choi, Y. (2020).
\newblock Defending {Against} {Neural} {Fake} {News}.
\newblock arXiv:1905.12616 [cs].

\bibitem[Zhao et~al., 2021]{zhao_deep_2021}
Zhao, B., Zhang, S., Xu, C., Sun, Y., and Deng, C. (2021).
\newblock Deep fake geography? {When} geospatial data encounter {Artificial}
  {Intelligence}.
\newblock {\em Cartography and Geographic Information Science}, 48(4):338--352.
\newblock Publisher: Taylor \& Francis \_eprint:
  https://doi.org/10.1080/15230406.2021.1910075.

\bibitem[Zhou and Leng, 2021]{zhou_china_2021}
Zhou, C. and Leng, S. (2021).
\newblock China mulls allowing more overseas investment in stocks, insurance.
\newblock Section: Economy.

\bibitem[Zhu et~al., 2023]{zhu_how_2023}
Zhu, J., Tian, Y.~L., and Tham, E. (2023).
\newblock How {China}'s new {No}.2 hastened the end of {Xi}'s zero-{COVID}
  policy.
\newblock {\em Reuters}.

\bibitem[Zucchi, 2021]{zucchi_technical_2021}
Zucchi, K. (2021).
\newblock Technical {Vs}. {Fundamental}: {Friends} or {Foes}?

\end{thebibliography}





\pagebreak
\onecolumn
\begin{center}
\textbf{\large Appendix A: Survey Design Structure}
\end{center}

\textbf{\large Survey in Chinese (Format of Delivery)}\newline
\begin{CJK*}{UTF8}{gbsn}
    \newline
    \noindent \textbf{说明}\newline
    
    \noindent 本调查旨在研究不同类型的公告和新闻对读者的影响。请仔细回答以下问题，在回答问题时不要依赖任何外部材料（例如使用互联网进行研究、询问朋友等）。完成调查的预计时间：阅读需要2分钟，回答问题需要5分钟。请不要在回答问题后再改答案。\newline
    
    \noindent \textbf{第一部分：预处理基线问题}\newline
    
    \begin{enumerate}
        \item 可口可乐公司是一家在股票市场上挂牌的国际饮料公司。您认为可口可乐公司2018年的\textbf{短期}经济前景如何?
        \begin{enumerate}
            \item 非常不利 / 有点不利 / 有点有利 / 非常有利
        \end{enumerate}
        \item 您认为可口可乐公司的\textbf{长期}经济前景如何？
            \begin{enumerate}
                \item 非常不利 / 有点不利 / 有点有利 / 非常有利 \newline
            \end{enumerate}
    \end{enumerate}
    
    \noindent \textbf{第二部分：处理}\newline
    
    \noindent 请阅读2018年可口可乐公司CEO发布的以下公告：\newline
\end{CJK*}    

\noindent [TREATMENT TEXT]\footnote{Treatment text and company names used within the survey depended on the randomized survey version that the participant received.}\newline
\begin{CJK*}{UTF8}{gbsn}
    \newline
    \noindent \textbf{第三部分：后续问题}\newline
    
    \begin{enumerate}
        \item 假设可口可乐公司的股票价格在这项公告之前为每股100美元。根据您对此公告的印象，可口可乐公司的每股股价变化了多少？
        \begin{enumerate}
            \item 下跌超过2美元 / 下跌在0.01美元至1.99美元之间 / 价格不变 / 上涨在0.01美元至1.99美元之间 / 上涨超过2美元
        \end{enumerate}
        \item 想象一下，如果在此公告发布时你得到了1000美元，你可以选择将其投资于可口可乐公司，或者保留现金。根据你对此公告的印象，你会投资多少钱在可口可乐公司?
        \begin{enumerate}
            \item \$0  / \$1 - \$250 / \$251 - \$500 / \$501 - \$750 / \$751 - \$1000
        \end{enumerate}
        \item 您认为该公司2018年的\textbf{短期}经济前景如何？
        \begin{enumerate}
            \item 非常不利 / 有点不利 / 有点有利 / 非常有利
        \end{enumerate}
        \item 请用一句话解释您回答问题3的考量。
        \begin{enumerate}
            \item 用选项回答。
        \end{enumerate}
        \item 请说明该公告是否完全不可信、不太可信、有些可信或非常可信。
        \begin{enumerate}
            \item 完全不可信 / 不太可信 / 有些可信 / 非常可信
        \end{enumerate}
        \item 请用一句话解释您回答问题5的考量。
        \begin{enumerate}
            \item 用选项回答。
        \end{enumerate}
        \item 您是否曾在金融服务业工作过全职工作？
        \begin{enumerate}
            \item 是 / 否
        \end{enumerate}
        \item 您是否曾经个人或通过专业服务提供商投资于在证券交易所上市的公司、指数或衍生品？
        \begin{enumerate}
            \item 是 / 否
        \end{enumerate}
        \item 您平均多久阅读一次商业新闻或金融文件？
        \begin{enumerate}
            \item 每天 / 每周 / 每月 / 每年 / 非常不经常或从不
        \end{enumerate}
        \item 您大约将您的净资产的百分之几投资于在证券交易所上市的公司、指数或衍生品中？
        \begin{enumerate}
            \item 0\% / 1\% - 25\%  / 26\% - 50\% / 51\% - 75\% / 76\% -100\%
        \end{enumerate}
        \item 你所获得的最高学历是什么？
        \begin{enumerate}
            \item 高中以下/ 高中毕业 / 大学学位/ 研究生学位（硕士、工商管理硕士、医学博士、法律博士、哲学博士等）
        \end{enumerate}
        \item 回顾过去一年，你的年收入是多少？
        \begin{enumerate}
            \item 少于50,000元人民币/ 介于50,000元人民币至99,999元人民币之间/ 介于100,000元人民币至149,999元人民币之间/ 介于150,000元人民币至199,999元人民币之间/ 大于199,999元人民币 / 我不要回答
        \end{enumerate}
        \item 你所属的种族或族裔群体是哪个？
        \begin{enumerate}
            \item 白人 / 非裔美国人或非洲裔美国人 / 亚洲或亚裔美国人 / 中东人 / 其他
        \end{enumerate}
        \item 无论问题是什么，请从以下选项中选择“我不知道”：“据您所知，美国的人口有多少？”
        \begin{enumerate}
            \item 5,000 万 / 3 亿 / 9 亿 / 10 亿 / 我不知道
        \end{enumerate}
        \item 你出生的年份是什么？
        \begin{enumerate}
            \item 在1964年之前 / 在1964年和1983年之间 / 在1984年和2003年之间 / 在2003年之后
        \end{enumerate}
        \item 你的性别是什么？
        \begin{enumerate}
            \item 男 / 女 / 其他
        \end{enumerate}
        \item 你的籍贯是哪个省份？
        \begin{enumerate}
            \item 省份（非中国籍）
        \end{enumerate}
    \end{enumerate}
\end{CJK*}

\textbf{\large Survey in English (Translation)}

\noindent \textbf{Instructions}\newline

\noindent This survey seeks to examine the effect of different kinds of announcements and news on readers. Please carefully answer the following questions, and do not rely on any external materials (e.g. using the internet to research, asking a friend, etc.) while you are answering. Estimated time for survey completion: 2 minutes for announcement reading, 5 minutes for question answering. Please do not change answers once you have completed a question. \newline

\noindent \textbf{Section 1: Pre-Treatment Questions}\newline

\begin{enumerate}
    \item The Coca-Cola Company is an international beverage company listed on the stock market. What do you think The Coca-Cola Company's \textbf{short-term} economic outlook was in 2018?
    \begin{enumerate}
        \item Very unfavorable / Somewhat unfavorable / Somewhat favorable / Very favorable
    \end{enumerate}
    \item What do you think was The Coca-Cola Company's \textbf{long-term} economic outlook?
        \begin{enumerate}
            \item Very unfavorable / Somewhat unfavorable / Somewhat favorable / Very favorable
        \end{enumerate}
\end{enumerate}

\noindent \textbf{Section 2: Treatment}\newline

\noindent Please read the following announcement made by The Coca-Cola Company's CEO in 2018: \newline

\noindent [TREATMENT TEXT]\footnote{Treatment text and company names used within the survey depended on the randomized survey version that the participant received.}\newline

\noindent \textbf{Section 3: Post-Treatment Questions}\newline

\begin{enumerate}
    \item Suppose that the stock price of The Coca-Cola Company was \$100 per share prior to this announcement. Based on your impression of this announcement, how much will the price per share of the Coca-Cola Company change?
    \begin{enumerate}
        \item Decrease more than \$2 / Decrease between \$0.01 – \$1.99 / No change in price / Increase between \$0.01 – \$1.99 / Increase more than \$2
    \end{enumerate}
    \item Imagine that you were given \$1000 USD at the time of this announcement. You can either invest in The Coca-Cola Company, or keep the money as cash. Based on your opinion of this announcement, how much of the \$1000 would you invest in The Coca-Cola Company?
    \begin{enumerate}
        \item \$0  / \$1 - \$250 / \$251 - \$500 / \$501 - \$750 / \$751 - \$1000
    \end{enumerate}
    \item What is your opinion on the \textbf{short-term} economic outlook of this company?
    \begin{enumerate}
        \item Very unfavorable / Somewhat unfavorable / Somewhat favorable / Very favorable
    \end{enumerate}
    \item In a few words, please explain your answer to question 3.
    \begin{enumerate}
        \item Write-in response
    \end{enumerate}
    \item Please describe whether the announcement was not at all believable, somewhat unbelievable, somewhat believable, or very believable.
    \begin{enumerate}
        \item Very unbelievable / Somewhat unbelievable / Somewhat believable / Very believable
    \end{enumerate}
    \item In a few words, please explain your answer to question 5.
    \begin{enumerate}
        \item Write-in response
    \end{enumerate}
    \item Have you ever worked a full-time job in the financial services industry?
    \begin{enumerate}
        \item Yes / No
    \end{enumerate}
    \item Have you ever invested in companies, indices, or derivatives listed on stock market exchanges, either personally or through a professional service provider?
    \begin{enumerate}
        \item Yes / No
    \end{enumerate}
    \item On average, how frequently do you read business news or financial documents? 
    \begin{enumerate}
        \item On a daily basis / On a weekly basis / On a monthly basis / On a yearly basis / Very infrequently or never
    \end{enumerate}
    \item Approximately what percentage of your net worth do you invest in companies, indices, or derivatives listed on stock market exchanges?
    \begin{enumerate}
        \item 0\% / 1\% - 25\%  / 26\% - 50\% / 51\% - 75\% / 76\% -100\%
    \end{enumerate}
    \item What is the highest level of education you have completed?
    \begin{enumerate}
        \item Primary school / Secondary school / College or University / Graduate program
    \end{enumerate}
    \item Thinking back on the last year, what was your annual income?
    \begin{enumerate}
        \item Less than 50,000 RMB / Between 50,000 RMB and 99,999 RMB / Between 100,000 RMB and 149,999 RMB / Between 150,000 RMB and 199,999 RMB / More than 199,999 RMB / I don't want to answer
    \end{enumerate}
    \item What racial or ethnic group best describes you?
    \begin{enumerate}
        \item White / African or African-American / Asian or Asian-American / Middle Eastern / Other
    \end{enumerate}
    \item Please select “I don’t know” from the options below regardless of the question: “To the best of your knowledge, what is the population of the United States?”
    \begin{enumerate}
        \item 50 million / 300 million / 900 million / 1 billion / I don't know
    \end{enumerate}
    \item What year were you born?
    \begin{enumerate}
        \item Before 1964 / Between 1964 and 1983 / Between 1984 and 2003 / After 2003
    \end{enumerate}
    \item What is your gender identification?
    \begin{enumerate}
        \item Male / Female / Other
    \end{enumerate}
    \item Where are you originally from?
    \begin{enumerate}
        \item Drop-down menu
    \end{enumerate}
\end{enumerate}

\pagebreak
\begin{center}
\textbf{\large Appendix B: Treatment Texts}
\end{center}

\textbf{\large Genuine Coca-Cola Transcript (Control)}

\textbf{Chinese Version}

\noindent \textit{Translation of English transcript was undertaken by two native Chinese speakers familiar with financial terminology.}\newline

\begin{CJK*}{UTF8}{gbsn}

\noindent 首先，我要强调我们正在一个充满活力的行业中取胜。全球饮料行业的增速超过去年，这背后主要的驱动力是新兴市场和发展中市场的业绩增长，和气泡类软饮销量上升。通过执行2017年初为制定的战略，我们获得了市场份额，经营业绩也因此而加速。\newline

\noindent 因此，在2018 年中的这个节点，我们看到了基础业务的良好势头。受销量和价格的双重推动，主体生意收入迄今增长 5\%，全球销量增长 3\%，这是我们五年来最强劲的增长。尽管汇率损失比我们在 2018 年初预期的要糟糕，公司每股收益EPS迄今已增长 5\%，而且正如在年初指出的那样，我们预计每股收益EPS 增长在后半年会更强劲。\newline

\noindent 在运营方面，放眼全球，我们在国际市场持续强劲的推动下，今年上半年实现了强劲的收入增长。在亚太地区，中国、印度和越南等主要新兴市场的强劲表现推动了该细分市场今年上半年实现4\%-6\%的有机收入增长。在同时扩大所有亚洲市场的饮料产品组合的同时，起泡软饮料产品组合加速了收入增长。\newline

\noindent 谈到拉丁美洲，即使面对阻力，我们在上半年也实现了高个位数的主体生意收入增长，因为我们专注于基本面。在起泡酒和乳制品产品组合稳健增长的推动下，墨西哥继续表现良好。在巴西，5 月下旬举行的为期 10 天的交通罢工实际上导致经济停滞，在那之前，本有望在第三季度实现连续的销量增长。通过系统性的努力来恢复业务增长，到本季度末，我们已经恢复到罢工前的增长率。\newline
\end{CJK*}

\textbf{English Version}

\noindent \textit{Transcript pulled from The Motley Fool's website (Coca-Cola Company (The) (KO) Q2 2018 Earnings Conference Call Transcript, 2018).}\newline

\noindent Let me start by saying and underlining that we're winning in a dynamic and vibrant industry. The global beverage industry is growing faster than last year, driven by better results in the emerging and developing markets, and also the sparkling soft drink category. We're gaining value share as we execute on the strategies we laid out for you in early 2017, and importantly, our operating performance is accelerating as a result.\newline

\noindent So, 2018 -- at the halfway mark of the year, we're seeing good momentum in our underlying business. Organic revenue is up 5\% year to date, driven by a good mixture of volume and price mix, global volume is up 3\%, which is the strongest we've seen in five years, and comparable EPS is up 5\% year to date despite stronger currency headwinds than we anticipated at the beginning of 2018, and, as we noted at the start of the year, we've always expected EPS growth to be back-half weighted, and that has not changed.\newline

\noindent So, operationally, looking around the world, we delivered strong top-line growth in the first half of the year, driven by continued strength in our international markets. In Asia-Pacific, strong performance in key emerging markets like China, India, and Vietnam drove mid-single-digit organic revenue growth for this segment during the first half of the year. While we continue to expand our total beverage portfolio across all of our Asia markets, our sparkling soft drink portfolio led our top-line acceleration.\newline

\noindent Turning to Latin America, even in the face of headwinds, we delivered high single-digit organic revenue growth for the first half as we focused on the fundamentals. Mexico continues to perform well, driven by solid growth across our sparkling and dairy portfolios. In Brazil, a ten-day transport strike in late May put a virtual halt to the economy. Prior to that, we were on track to deliver the third quarter of sequentially improving volume growth. Our system worked hard to recover momentum, and by the end of the quarter, we had bounced back to pre-strike growth rates.\newline

\textbf{\large Synthetic Coca-Cola Transcript}

\textbf{Chinese Version}

\noindent \textit{Prompt: "Translate the following
into Chinese: [Treatment Text]"} \newline

\begin{CJK*}{UTF8}{gbsn}
    
    \noindent 大家早上好，谢谢大家。在介绍我们第二季度业绩之前，我想花一点时间来谈一下当前的经济环境。我们正面临着一个充满挑战和不确定性的全球经济形势，波动性和风险越来越大。\newline
    
    \noindent 这种环境正在影响消费者信心和购买行为，尤其是在新兴市场，我们在最近几个季度中看到了明显的放缓。我们也面临着不断上涨的投入成本，特别是像铝这样的商品，这给我们的利润带来了压力。\newline
    
    \noindent 鉴于这些不利因素，我们正在谨慎和有纪律地管理我们的业务，同时继续投资于我们的品牌和创新项目。我们有信心可以长期应对这些挑战并更强大地崛起。\newline
    
    \noindent 现在，转向我们的第二季度业绩，我们实现了89亿美元的净营收，同比下降2\%。这一下降是由多种因素造成的，包括不利的货币波动。\newline
    
    \noindent 我们的营业利润下降了5\%，主要是由于投入成本上升和持续投资于我们的增长项目。我们专注于管理成本，并改善我们业务的生产力以缓解这些压力。\newline
    
    \noindent 就我们的类别表现而言，碳酸软饮料下降了1\%，主要是由于拉丁美洲和欧洲的下降。然而，我们在北美和亚洲看到了可口可乐零糖和健怡可乐品牌推动的强劲增长。\newline
    
    \noindent 我们的非气泡饮料组合下降了2\%，主要是由于果汁和果汁饮料的下降，部分抵消了茶和咖啡的增长。我们仍然看到我们的高端水品牌，包括Smartwater和Topo Chico的强劲增长。\newline
    
    \noindent 就我们的全球市场份额而言，我们保持了领先的非酒精饮料公司的地位，占据总市场的19\%。然而，我们正面临来自传统和非传统参与者的增加的竞争，尤其是在健康和生活方式领域。\newline
\end{CJK*}

\textbf{English Version}

\noindent \textit{Prompt: "Generate the first 550 words of a fictional Q2 2018 earnings call transcript for The Coca-Cola Company with a negative economic outlook."}\newline

\noindent Thank you, and good morning, everyone. Before we get into our second-quarter results, I want to take a moment to address the current economic environment. We are facing a challenging and uncertain global economic landscape, with increasing volatility and risks.\newline

\noindent This environment is impacting consumer sentiment and purchasing behavior, particularly in emerging markets, where we have seen significant slowdowns in recent quarters. We are also facing rising input costs, particularly in commodities like aluminum, which is putting pressure on our margins.\newline

\noindent Given these headwinds, we are taking a prudent and disciplined approach to managing our business, while continuing to invest in our brands and innovation pipeline. We are confident in our ability to navigate these challenges and emerge stronger in the long run.\newline

\noindent Now, turning to our second-quarter results, we delivered net revenue of \$8.9 billion, a decrease of 2\% year over year. This decline was driven by a combination of factors, including unfavorable currency movements.\newline

\noindent Our operating income declined by 5\%, primarily due to higher input costs and continued investments in our growth initiatives. We are focused on managing our costs and improving productivity across our operations to mitigate these pressures.\newline

\noindent Looking at our category performance, sparkling soft drinks were down 1\%, led by declines in Latin America and Europe. However, we saw strong growth in North America and Asia, driven by our Coca-Cola Zero Sugar and Diet Coke brands.\newline

\noindent Our still beverage portfolio declined by 2\%, primarily due to declines in juice and juice drinks, partially offset by growth in tea and coffee. We continue to see strong growth in our premium water brands, including Smartwater and Topo Chico.\newline

\noindent In terms of our global market share, we maintained our position as the leading non-alcoholic beverage company, with a 19\% share of the total market. However, we are facing increased competition from both traditional and non-traditional players, particularly in the health and wellness space.\newline

\textbf{\large Synthetic HP Transcript}

\textbf{Chinese Version}

\noindent \textit{Prompt: "Translate the following
into Chinese: [TREATMENT TEXT]"}\newline

\begin{CJK*}{UTF8}{gbsn}

    \noindent 下午好，欢迎参加惠普公司2018财年第二季度财报电话会议。今天，我们将讨论截至2018年4月30日季度的财务结果。不幸的是，我必须承认，在这个季度中，我们面临了重大挑战，我们的业绩反映了我们所处的艰难经济环境。\newline
    
    \noindent 首先，让我简要介绍一下本季度的财务结果。我们报告了139.3亿美元的收入，同比下降9\%。我们的摊薄每股收益为0.48美元，同比下降14\%。我们的非通用会计准则摊薄每股收益为0.55美元，同比下降7\%。\newline
    
    \noindent 正如您所看到的，我们在本季度面临了多个领域的阻力。我们的个人系统部门特别受到消费者市场的影响，导致我们的笔记本电脑和台式机销售低于预期。此外，我们在消费和商业市场都看到了打印机和耗材的需求疲软。我们的耗材业务的下降尤其具有挑战性，因为它一直是惠普的高利润和稳定的收入来源。\newline
    
    \noindent 展望未来，我们预计经济环境在财年剩余时间内仍将面临挑战。特别是，我们预计消费者市场将继续疲软，这可能会影响我们的个人系统业务。此外，我们预计我们的耗材业务将继续面临压力，因为客户寻求更低成本的替代方案或推迟购买。\newline
    
    \noindent 尽管面临这些挑战，我们仍然致力于我们的长期战略和我们对创新和客户体验的关注。我们继续投资于我们的产品组合和我们的市场推广能力，以便让惠普在未来取得成功。\newline

\end{CJK*}

\textbf{English Version}

\noindent \textit{Prompt: "Generate the first 550 words of a fictional Q2 2018 earnings call transcript for HP Inc. with a negative economic outlook. Restrict output to comments by the CEO."}\newline

\noindent Good afternoon, and welcome to HP Inc.'s Q2 2018 earnings call. Today, we will be discussing our financial results for the quarter ended April 30, 2018. Unfortunately, I must start by acknowledging that we faced significant challenges during the quarter, and our results reflect the difficult economic environment we are operating in.\newline

\noindent First, let me give you a brief overview of our financial results for the quarter. We reported revenue of \$13.93 billion, which is down 9\% year over year. Our GAAP diluted earnings per share were \$0.48, which is down 14\% year over year. Non-GAAP diluted earnings per share were \$0.55, which is down 7\% year over year.\newline

\noindent As you can see, we faced headwinds in multiple areas during the quarter. Our Personal Systems segment was particularly impacted by a weak consumer market, which led to lower-than-expected sales of our notebooks and desktops. Additionally, we saw soft demand for our printers and supplies in both the consumer and commercial markets. The decline in our Supplies business was particularly challenging, as it has historically been a high-margin and stable source of revenue for HP.\newline

\noindent Looking forward, we expect the economic environment to remain challenging for the remainder of the fiscal year. In particular, we anticipate continued weakness in the consumer market, which will likely impact our Personal Systems business. Additionally, we expect ongoing pressure in our Supplies business as customers seek lower-cost alternatives or delay purchases.\newline

\noindent Despite these challenges, we remain committed to our long-term strategy and our focus on innovation and customer experience. We continue to invest in our product portfolio and our go-to-market capabilities to position HP for success in the future.\newline

\pagebreak
\begin{center}
\textbf{\large Appendix C: Metric Accuracy Within Synthetic Coca-Cola Treatment Text}
\end{center}

\noindent Thank you, and good morning, everyone. Before we get into our second-quarter results, I want to take a moment to address the current economic environment. We are facing a challenging and uncertain global economic landscape, with increasing volatility and risks.\newline

\noindent This environment is impacting consumer sentiment and purchasing behavior, particularly in emerging markets, where we have seen significant slowdowns in recent quarters. We are also facing rising input costs, particularly in commodities like aluminum, which is putting pressure on our margins.\newline

\noindent Given these headwinds, we are taking a prudent and disciplined approach to managing our business, while continuing to invest in our brands and innovation pipeline. We are confident in our ability to navigate these challenges and emerge stronger in the long run.\newline

\sethlcolor{green}

\noindent Now, turning to our second-quarter results, we delivered \hl{net revenue of \$8.9 billion},\footnote{Correct (Q2 2018 Earnings Release, 2018).} \sethlcolor{magenta} a \hl{decrease of 2\% year over year}.\footnote{Incorrect. Actual metric: -8\% (Q2 2018 Earnings Release, 2018).} This decline was driven by a combination of factors, including unfavorable currency movements.\newline

\noindent Our \hl{operating income declined by 5\%},\footnote{Incorrect. Actual metric: +34\% (Q2 2018 Earnings Release, 2018).} primarily due to higher input costs and continued investments in our growth initiatives. We are focused on managing our costs and improving productivity across our operations to mitigate these pressures.\newline

\noindent Looking at our category performance, sparkling soft drinks were \hl{down 1\%},\footnote{Incorrect. Actual metric: +2\% (Q2 2018 Earnings Release, 2018).} led by declines in Latin America and Europe. However, we saw strong growth in North America and Asia, driven by our Coca-Cola Zero Sugar and Diet Coke brands.\newline

\sethlcolor{gray}

\noindent Our still beverage portfolio \hl{declined by 2\%},\footnote{Unverifiable.} primarily due to declines in juice and juice drinks, partially offset by growth in tea and coffee. We continue to see strong growth in our premium water brands, including Smartwater and Topo Chico.\newline

\noindent In terms of our global market share, we maintained our position as the leading non-alcoholic beverage company, with a \hl{19\% share of the total market}.\footnote{Unverifiable.} However, we are facing increased competition from both traditional and non-traditional players, particularly in the health and wellness space.\newline

\pagebreak
\begin{center}
\textbf{\large Acknowledgements}
\end{center}

 I want to first take this opportunity to formally express my gratitude towards the many individuals that helped make this project possible:\newline

    \begin{itemize}
    
        \item \noindent To the Schwarzman Scholars program -- for providing a once-in-a-lifetime opportunity to study the world from such a unique perspective.\newline
        
        \item \noindent To Dr. Zeng Yi -- for providing guidance, recommendations, and funding support for the creation of an empirical study through the Chinese Academy of Sciences -- as well as Zhang "Kang" Dabao, for helping with its translation and implementation in China.\newline
    
        \item \noindent To Flora Liao, Lin Ye, Xiran Cheng, Jojo Zhao, and Mikaela Wang -- for helping review and adjust the translations within this study.\newline
    
        \item \noindent To my friends and classmates at Schwarzman College -- for entertaining my musings and ramblings on this topic both inside and outside the classroom.\newline 
        
        \item \noindent And finally, to my family members -- who endured endless conversations about my interest in this area, as well as how to process my ideas.\newline
        
    \end{itemize}
    
    My concern surrounding synthetic media began around 2020/2021, when I began reading about the unprecedented bipartisan mistrust that exists towards mainstream media sources within American society. During the same time period, I was taking a class on ``Deep Learning." During one lecture, my class was presented with ``deepfake" satellite images that were indistinguishable from real ones (similar to the ones shown in Chapter 2). It was at that moment that I first began to (fearfully) consider the possibility of a future in which we -- as a society -- don't know what is true, or even real.

    Among other reasons, I came to the Schwarzman Scholars program in 2022 to take a first step towards defining how realistic this concern was, as well as whether it was a problem ``worth solving" in the world. In August of the same year, when I started formulating the ideas for this capstone, I wasn't sure I had chosen the right topic. After all, concerns about synthetic media seemed to have dissipated in the context of invalid election tampering fears in 2020, as well as the Zelensky video's quick removal from social media channels in March.

    The wave of discussion and controversy around synthetic content generation that came alongside ChatGPT's release in November of 2022 changed that. On the one hand, I feel relieved that the concern surrounding synthetic media's impact on society has re-risen to the forefront of discussion in the artificial intelligence community. On the other hand, I feel a growing sense of dread -- that perhaps global information environments are changing faster than we hope/can control.
    
    Regardless, I imagine it's rare to find oneself in a situation where historical research becomes current overnight, and I feel fortunate to have been able to witness and participate in that process. I hope that this study will serve as a baseline for future research, as well as for a better future.

    - Nick Keeley

\end{document}